\documentclass[reprint,pra,superscriptaddress,showpacs,twocolumn]{revtex4-2}
\usepackage{graphicx}
\usepackage[ansinew]{inputenc}
\usepackage{amsfonts}
\usepackage[normalem]{ulem}
\usepackage{amsmath}
\usepackage{mathrsfs}
\usepackage{color}
\begin{document}
\title{Enhanced solar photocurrent using a quantum-dot molecule}

\author{J. Lira}
\email{jefferson.santos@ufu.br}
\affiliation{Instituto de F\'{i}sica, Universidade Federal de Uberl\^andia,
38400-902, MG, Brazil}
\affiliation{Colegiado de Engenharia, UniFTC, 45020-510, Vit\'oria da Conquista-BA, Brazil}
\author{J. M. Villas-Boas}
\email{boas@ufu.br}
\affiliation{Instituto de F\'{i}sica, Universidade Federal de Uberl\^andia,
38400-902, MG, Brazil}
\author{L. Sanz}
\email{lsanz@ufu.br}
\affiliation{Instituto de F\'{i}sica, Universidade Federal  de Uberl\^andia,
38400-902, MG, Brazil}
\author{A. M. Alcalde}
\email{augusto.alcalde@ufu.br}
\affiliation{Instituto de F\'{i}sica, Universidade Federal de Uberl\^andia,
38400-902, MG, Brazil}

\date{\today}

\begin{abstract}
We present a detailed study on the influence of coherent tunneling on the photovoltaic properties of a semiconductor molecule driven by solar radiation. The connection between the power delivered by the QDM and quantum coherence is not simply proportional but depends on an interplay between the interdot coherent tunneling, the interaction of the system with thermal phonon reservoirs, and the resonance between the QDM and conduction bands. We explored numerically various parameter regimes and found that the maximum power delivered by the molecule is up to 30\% greater than the power delivered by a single quantum-dot device. The calculated photovoltaic conversion efficiency is presented in terms of accessible experimental parameters and, as expected, is constrained by the second law.
\end{abstract}
\maketitle
\section{Introduction}
\label{sec:intro}
The efficiency of solid state solar cells is limited by various extrinsic factors that in principle can be minimized through careful design and materials engineering. In parallel, one of the intrinsic factors that determine the behavior of photovoltaic efficiency is radiative recombination of the electron-hole pairs before they can be extracted during photovoltaic conversion. In the classic context of detailed balance, the radiative recombination balances radiative absorption and thus establishes a fundamental limit to the efficiency of photovoltaic conversion~\cite{Shockley1961}. More recently, it has been proposed that interference effects resulting from quantum coherence could be used to enhance the photovoltaic conversion efficiency in photocells~\cite{Zhao2019,Dorfman2018,Creatore2013,Scully2010,Dorfman2011,Svidzinsky2011}. These proposals tackle the problem from different perspectives such as the use of the noise-induced coherence~\cite{Dorfman2018}, or the interference of photon absorption and emission induced by the dipole-dipole interaction between molecular excited states~\cite{Creatore2013}. Other approaches include considering an external driving field to induce coherence between two levels in a photocell~\cite{Scully2010,Dorfman2011}, or quantum interference mechanisms such as Fano coupling~\cite{Svidzinsky2011}.

In recent decades, semiconductor quantum dots (QDs) have been used as active nanostructured absorbers in solar cell technology~\cite{Aroutiounian2012,Kamat2013},  with different degrees of success~\cite{Guimard2010,Beattie2017,Pan2013,Santra2012,Cirloganu2014,Lee2013}. Most efforts have been directed to the development of new materials~\cite{Polman2016} with favorable optical properties, band optimization~\cite{Luque1997,Okada2015}, and efficiency in the charge transfer mechanisms~\cite{Bang2009}. Continued progress in nanofabrication also allowed a fine adjustment of electron and hole band offsets~\cite{Bracker2006}, the high controllability of coherent effects in quantum dot systems~\cite{Ramsay2010}, as well as the control of optical excitations in tunneling coupled quantum dots~\cite{Kim2011,Greilich2011}, and the variety of quantum interference phenomena related to coherent population trapping~\cite{Weiss2012,Borges2012,Borges2013}. Photovoltaic conversion in arrays of self-assembled QDs have recently been experimentally studied, showing an efficiency of 18.3\% at 5 suns concentration~\cite{Beattie2017}.

In this paper, we study the contribution of coherence, generated by the tunneling processes of electrons and holes, on the photocurrent and the delivered power by a III-V quantum dot molecule (QDM) embedded in a photodiode structure. The dynamics of the excited states produced by incoherent natural thermal light is analyzed based on the framework of the Weisskopf-Wigner approach~\cite{Scully2010,Svidzinsky2011}. Taking advantage of the high structural controllability of the  nanostructure, we first explore the photocurrent and power delivered by the QDM considering different scenarios of barrier thickness and radiative recombination decays. Then, we connect our findings with the behavior of the quantum coherence under several choices of the physical parameters. We also calculate the photovoltaic properties considering the competitive effect between coherent mechanisms and different configurations of band alignment. We demonstrate that the photocurrent for our QDM is in a range of 8\% and 30\%. These results, together with technics that allow the fabrication of ensembles of nanostructures with a high dot density value~\cite{Chia07}, had the potential to raise the net gain on photovoltaic conversion.

This paper is organized as follows. In Sec.~\ref{sec:model} we present the model of the QDM, and relevant physical parameters. The analysis of the effect of coherent tunneling in the production of photocurrent is discussed in Sec.~\ref{sec:cohtunnel}. Section~\ref{sec:bands} is devoted to the study of the effect of the alignments (detunings) of electrodes and electron and holes energy levels, and a brief discussion about efficiency. Finally, Sec.~\ref{sec:summary} contains a summary of our work.

\section{The quantum-dot molecule as a photocurrent source}
\label{sec:model}
\subsection{Physical System and Model}
\begin{figure}[t]
	\centering\includegraphics[scale=1.3]{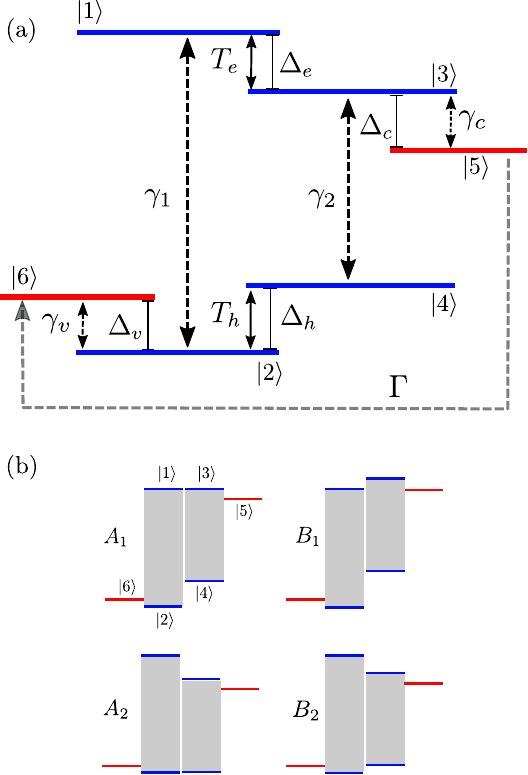}
	\caption{(a) Schematic six-level model of the QDM. The blue lines represent the energy levels of QDM and the red lines the conduction and valence band states of contact electrodes. Solar radiation continuously excite the electron transitions $\vert 2 \rangle \leftrightarrow \vert 1 \rangle $ and $\vert 4 \rangle \leftrightarrow \vert 3 \rangle $. Ambient thermal phonons mediate the low energy transitions $\vert 3 \rangle \leftrightarrow \vert 5 \rangle$ and $\vert 2 \rangle \leftrightarrow \vert 6 \rangle $ at rates $\gamma_{c}$ and $\gamma_{v}$, respectively. Contact levels $\vert 5 \rangle$ and $\vert 6 \rangle$ are connected to a load (external circuit, represented by the light-gray dashed line), with decay rate $\Gamma$. $\Delta_{e,h}$ are the conduction and valence band offsets of QDM, while $\Delta_{c,v}$ are related with the electrodes. (b) Resonant energy alignments where in $A_1$ ($A_2$) the conduction (valence) levels of the QDM are resonant, whereas in $B_2$ ($B_2$) the conduction (valence) level of the left (right) QD is resonant with the conduction (valence) of the contact.}
	\label{fig1}
\end{figure}

We consider a typical InAs/GaAs semiconductor QDM, composed of two vertically aligned QDs of different sizes, and separated by a tunneling barrier of width $d$. Initially the QDM has no charge and, after optical excitation, excited electrons and holes are able to tunnel between the quantum dots~\cite{Bracker2006}. The energy levels labeled as  $\vert 1 \rangle$, $\vert 3 \rangle$  ($\vert 2 \rangle$, $\vert 4 \rangle$), indicated by blue lines in the figure~\ref{fig1}, represent the electron (hole) states of the QDM. Inside the QDM, the conduction levels $\vert 1 \rangle$ and $\vert 3 \rangle$ are coupled by tunneling $T_e$, while the tunneling between the QDM valence states $\vert 2 \rangle$ and $\vert 4 \rangle $ is given by $T_h$. 

Apart from the coherent tunneling, the model includes incoherent processes. First, we treat the sun radiation as a hot reservoir, which drives the electron-hole transitions $\vert 2 \rangle \leftrightarrow \vert 1 \rangle$ and $\vert 4 \rangle \leftrightarrow \vert 3 \rangle$, indicated by black dashed lines in Fig.~\ref{fig1}. We consider that the average photon occupations given by $n_1 = \left[ \exp(E_{12}/k_BT_S) -1 \right]^{-1}$, and $n_2 = \left[ \exp(E_{34} / k_BT_S) -1 \right]^{-1}$, respectively, where $E_{ij} = E_i - E_j$ is the QDM electron-hole transition energy, $T_S$ is the sun temperature, and $k_B$ is the Boltzmann constant. The electron-hole recombination rate is given by $\gamma_1$ $(\gamma_2)$ for $E_{12}$ $(E_{34})$ transitions. We model a second process, the coupling of the QDM with cold reservoirs, by considering the energy levels $\vert 5 \rangle $ and $ \vert 6 \rangle $ (red lines in Fig.~\ref{fig1}(a)), which represent the conduction and valence states of the photodiode in which the QDM is embedded, respectively. We assume that ambient thermal phonons at temperature $T_c$ couple the low energy transitions $\vert 3 \rangle \leftrightarrow \vert 5 \rangle$ and $\vert 2 \rangle \leftrightarrow \vert 6 \rangle$, with average phonon occupations $n_c = \left[ \exp(E_{35}/k_BT_c) - 1\right]^{-1}$, and $n_v = \left[ \exp(E_{62}/k_BT_c) - 1\right]^{-1}$, respectively~\cite{Scully2010,Dorfman2011}. After the charge separation and collection, the conduction and valence reservoir states are connected to a lead. We model this process as a decay rate $\Gamma$ of the conduction level $\vert 5 \rangle \rightarrow \vert 6 \rangle$, illustrated by the light-gray dashed lines in Fig.~\ref{fig1}.

The hamiltonian which describes the model is written as
\begin{equation}
	\label{eq:fullH}
	\hat{H}=\hat{H}_{0}+\hat{H}_B+\hat{V}_{\mathrm{hot}}+\hat{V}_{\mathrm{cold}},
\end{equation}
where $\hat{H}_{0}$ is the free hamiltonian for the 6-level system written as
\begin{equation}
	\label{eq:Ho}
	\hat{H}_{0}=\hbar\sum_{i}^{}\omega_i \left| i\right\rangle \left\langle i\right|+T_e \left|1\right\rangle \left\langle 3\right| + T_h \left| 2\right\rangle \left\langle 4\right| + \mathrm{H.c.},
\end{equation}
with $i=1,...,6$ label the energetic levels on Fig.~\ref{fig1}, and the terms depending of $T_{e}$ and $T_{h}$ describe the electron and hole coherent tunnelings respectively. In order to include QDM barrier thickness $d$ dependence on tunneling coupling, we fit the experimental data from Ref.~\cite{Bracker2006} for an {InAa/Gas} QDM with exponentially decreasing functions of type $T_{e,h} \sim e^{-d/d_{e,h}}$~\footnote{From data of Ref. \cite{Bracker2006}, the electron and hole anticrossings energies, $\Delta_{e,h}$, due to tunneling coupling, decrease exponentially with $d$. We found by fitting $\Delta_e \simeq 11.67 e^{-d/7.14}$ and $\Delta_h \simeq 2.2 e^{-d/3.37}$, respectively. Using a simple two-level model the corresponding tunneling rates are given by $T_i = \Delta_i /2\hbar$, with $i=e,h$.}, where $d_e = 7.14$~nm and $d_h = 3.37$~nm.

The second term in Eq.(\ref{eq:fullH}), $\hat{H}_B$, is the free Hamiltonian for the reservoirs written as
\begin{equation}
	\hat{H}_B=\sum_{k}\hbar\nu_k\hat{a}^{\dagger}_k\hat{a}_k + \sum_{l}^{}\hbar\nu_l\hat{c}^{\dagger}_l\hat{c}_l + \sum_{m}^{}\hbar\nu_m\hat{v}^{\dagger}_m\hat{v}_m,
\end{equation}
and the last two terms describe the interaction between the system and the reservoirs, being
\begin{eqnarray}
	\hat{V}_{\mathrm{hot}}&=& \hbar\sum_{k}^{}g_{k}\left(\left|1\right\rangle\left\langle 2\right|+\left|3\right\rangle\left\langle 4\right|\right)\hat{a}_k + \mathrm{H.c.},\\\nonumber
	\hat{V}_{\mathrm{cold}}&=& \hbar\sum_{l}^{}g_{l}\left|3\right\rangle\left\langle 5\right|\hat{c}_l + \hbar\sum_{m}g_{m}\left|6\right\rangle\left\langle 2\right|\hat{v}_m+ \mathrm{H.c.}.
\end{eqnarray}
In this three equations the hot reservoir corresponds to the solar radiation with frequency $\nu_k$, and $\hat{a}_k$ ($\hat{a}^{\dagger}_k$) being the annihilation (creation) operator. The cold reservoirs are two different baths, associated with phonons from the conduction and valence bands of the host semiconductor in which the QDM is embedded, where the operators $\hat{c}_l$ ($\hat{c}^{\dagger}_l$) and $\hat{v}_m$ ($\hat{v}^{\dagger}_m$) are the annihilation (creation) operator describe the conduction and valence bands phononic reservoirs, with frequency $\nu_l$ and $\nu_m$ respectively. The constants $g_B$ with $B=k,l,m$ describes the coupling between system and each bath and we do not consider the polaronic effect in our treatment~\footnote{Although manipulations on the growth process could enhance the coupling between electrons and phonons for some very specific situations, the coupling can be considered weak, because the local charge inside the quantum dot is neutral (each dot has the same number of electrons and holes in the steady-state).}~\cite{Heitz01,Schmitt87}.

By tracing out the degrees of freedom of the reservoirs, it is straightforward to check that the evolution of this system is governed by the master equation
\begin{equation}
	\dot{\hat{\rho}} = -\frac{i}{\hbar}[\hat{H}_{0}, \hat{\rho}] + \mathcal{L}_{\mathrm{hot}}[\hat{\rho}] + \mathcal{L}_{\mathrm{cold}}[\hat{\rho}].
	\label{eq:Lindblad}
\end{equation}
Here $\mathcal{L}_{\mathrm{hot}}[\hat{\rho}]$ represents the Liouvillian which describes the interaction between QDM and the sun radiation, which is written as
\begin{eqnarray}
	\label{eq:liovillian}
	\mathcal{L}_{\mathrm{hot}}[\hat{\rho}]&=&\frac{1}{2}\sum_{s=a,b}^{}\{\gamma_s(n_s + 1)([\sigma_s\hat{\rho},\sigma_s^{\dagger}] + [\sigma_s,\hat{\rho}\sigma_s^{\dagger}])\nonumber\\
	&&+\gamma_s n_s ([\sigma_s^{\dagger}\hat{\rho},\sigma_s] + [\sigma_s^{\dagger},\hat{\rho}\sigma_s])\},
\end{eqnarray}
where $s=a$ label the transitions $\left|1\right\rangle \leftrightarrow\left| 2\right\rangle$ in the first quantum dot, associated with the transition operator $\sigma_a=\left|2\right\rangle \left\langle 1\right|$, decay rate $\gamma_1$ and average number of photons $n_1$. The label $s=b$ stands for the process $\left|3\right\rangle \leftrightarrow\left| 4\right\rangle$ (second quantum dot) with operator $\sigma_b=\left|4\right\rangle \left\langle 3\right|$, decay rate $\gamma_2$ and average number of photons $n_2$. In turn, the second term $\mathcal{L}_{\mathrm{cold}}[\hat{\rho}]$, representing the action of the phononic reservoirs, has a similar form that can be obtaining by changing the label $s$ for $p$, and considering $p=c,v$. Then, the label $p=c$ is reserved for the process $\left|3\right\rangle \leftrightarrow\left| 5\right\rangle$ relative to the QDM coupling of conduction band with operator $\sigma_c=\left|5\right\rangle \left\langle 3\right|$, decay rate  $\gamma_c$ and number average of phonons $n_c$. For $p=v$, the associated process is $\left|6\right\rangle \leftrightarrow\left| 2\right\rangle$ relative to the QD-contact coupling of the valence band with operator $\sigma_v=\left|2\right\rangle \left\langle 6\right|$, decay rate $\gamma_v$ and number average of phonons $n_v$. Here, the decay rates are given by~\cite{JauhoBook} $\gamma_{s(p)}=2\pi D\left(\omega_{s(p)}\right)\omega^2_{s(p)}g^2_{s(p)}/v^3$,
where $\omega^2_{s(p)}$ is the frequency of the transitions, $g_{s(p)}$ are the couplings, and $D\left(\omega_{s(p)}\right)$ are the density of states for each process $s=1,2$ and $p=c,v$. Once we are interested on the steady-state, we can assume this values as constants, as in several works~\cite{Scully2010,Svidzinsky2011}, using experimental measurements of the process of relaxation of the electron from the conduction to the valence bands. The Eq.~(\ref{eq:Lindblad}) consists on a system of coupled differential equations shown in details in the Appendix~\ref{app:coupleddeq}, Eqs.(\ref{eq:scde}). By solving this system of coupled equations in the stationary regime considering $\dot{\rho} = 0$, we obtain numerically the steady populations, $\rho_{ii}$, and coherences, $\rho_{ij}$.

\subsection{Characterization of quantum dots and molecules as photovoltage sources}
Once we are interested on the application of a QDM as a photovoltaic device, we present some definitions. The photocurrent $j$ is defined in terms of the population of state $\vert 5 \rangle$ being  
\begin{equation}
	j = e \Gamma \rho_{55}.
\end{equation}
The power $P$ delivered by the QDM is calculated as usual 
\begin{equation}
	P = jV.
\end{equation} 
Finally, the photovoltage across the device is given by 
\begin{equation}
	V = (E_{5} - E_6) + k_BT_c \ln(\rho_{55}/\rho_{66}).
\end{equation}
The efficiency at maximum power of the photovoltaic device, $\eta$ is calculated considering the maximum power delivered, $P_m$ as
\begin{equation}
	\label{eq:efficiency}
	\eta = P_m / P_S,
\end{equation} 
where $P_S$ is the power supplied by the incident solar radiation~\cite{Svidzinsky2011} which is defined as
\begin{equation}
	\label{eq:powersupplied}
	P_S = j E_{12}/e.
\end{equation}

Before apply our model considering the physical parameter of quantum molecules with coherent tunneling~\cite{Stinaff636}, we perform a quick comparison with an experiment on actual quantum dot solar cell (QDSC) reported by Guimard and co-workers~\cite{Guimard2010}. We consider first a single quantum dot in our model and adjust the value of the energy gap as $E_{12}=920$ meV. This value provide us an open-circuit voltage given by $V_{OC}=871$ mV, which coincide with the value reported by the authors for one of the QDSC in the reference. The electron-hole recombination rate on our model was set to $\gamma_1=0.19\gamma$ to fit the contribution of the quantum dot to the short-circuit current density\footnote{This value was obtained after extracting the experimental data from Fig. 2 of the external quantum efficiency reported in Guimard \textit{et al.} and additional calculations.}. After fitting, we obtain a short-circuit current value of $j_{55}/e\gamma=0.018$, with a maximum power of $P_m/\gamma=13.66$ meV. which is consistent with the reported values in Guimard \textit{et al.}~\cite{Guimard2010}. The results let us confident that our model is well succeeded in simulating a realistic QDSC. Extending the simulation to a situation where two identical quantum dots, coupled by tunneling, as described above are considered, we verify that considering a separation of $d=2$ nm between the QDs the short-circuit current increases to $j_{55}/e\gamma=0.0300$, with an also increase of the maximum power output of $P_m/\gamma=22.28$ meV, showing a gain in both, short-circuit current and power, associated with the effect of tunneling.

\section{The effect of coherent tunneling in photovoltaic properties of a QDM}
\label{sec:cohtunnel}
In this section, we present a study of the effect of the coherent tunneling on the properties of generation of photocurrent considering a QDM, which are experimentally realistic values for III-V semiconductor QDMs~\cite{Bracker2006,Stinaff636}. Our interest on this specific experimental plataform is justified on the demonstration of coherent tunneling~\cite{Stinaff636}. For a InAs/GaAs QDM, we consider $\Delta_e = \Delta_h = 3$ meV, $\Delta_c = \Delta_v = 2$ meV. Also, the energy-gap are given by $E_{12} = 1115$ meV and $E_{34} = E_{12} - (\Delta_e + \Delta_h)$. We chose tunneling barrier widths in the range $d = 2 - 10 $ nm. Within this range, the electron and hole tunneling rates are $\hbar T_e = 4.41-1.44$meV, and $\hbar T_h = 0.56-0.06$meV, respectively. Temperatures are set on $k_BT_S=500$ meV and $k_BT_c= 25.9$ meV. To find numerically the characteristic current-voltage relation, we keep all the parameters fixed while varying the value of $\Gamma$, from $\Gamma = 0$ corresponding to the open circuit, to $\Gamma \rightarrow \infty$, the short circuit limit~\cite{Dorfman2011}. To investigate the effect of the coherent tunneling on the photovoltaic properties of the system, we vary the separation between QDs $d$, which in turn changes the coherent tunneling couplings $T_e$ and $T_h$, as discussed above. 

\subsection{Electronic current and power}

\begin{figure}[t]
	\centering\includegraphics[scale=0.5]{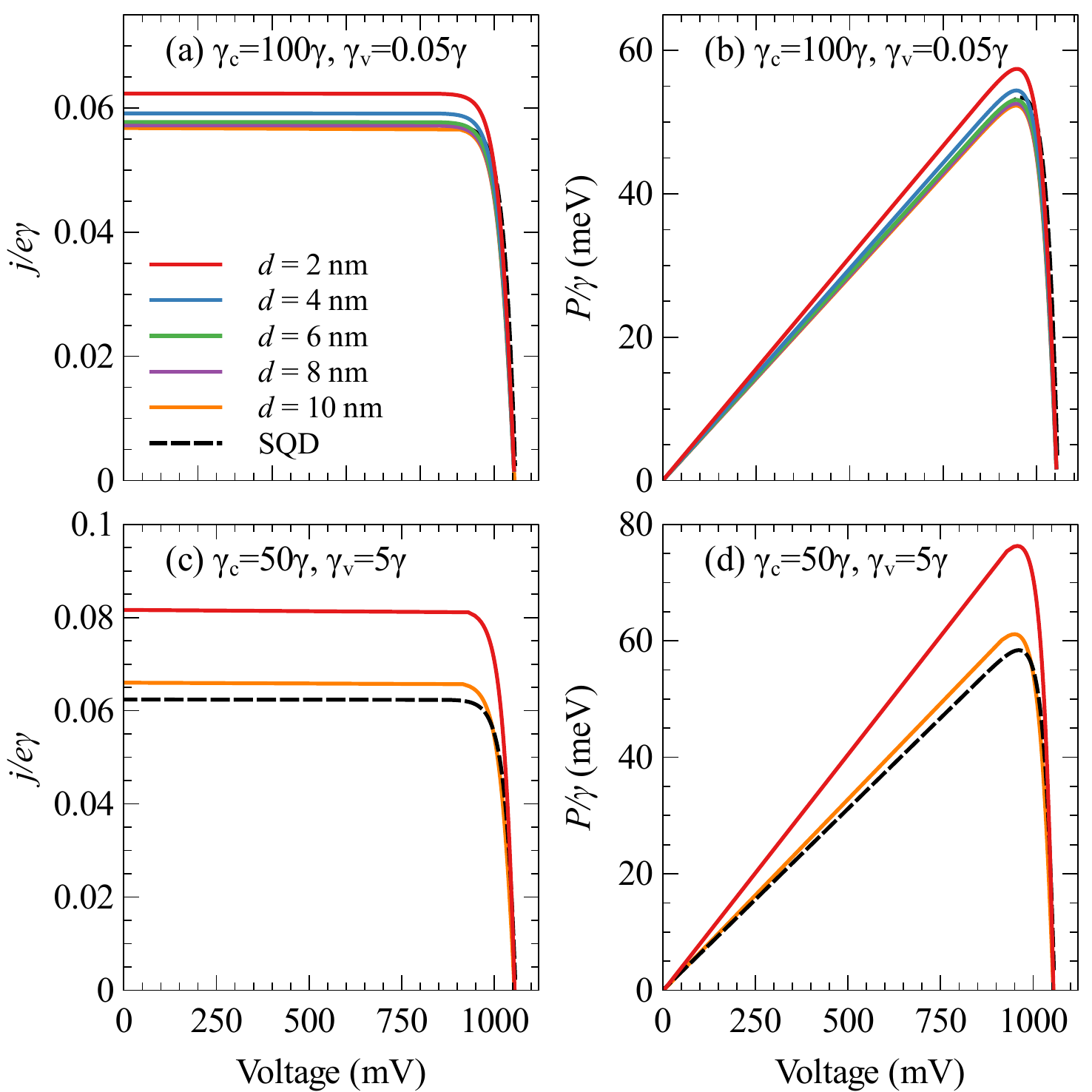}
	\caption{(Color online) (a) The QDM characteristic current-voltage relation and, (b) the power generated as functions of voltage considering the values $\gamma_{c} = 100\gamma$, $\gamma_{v} = 0.05\gamma$. Panels (c) and (d) shown the same physical quantities considering a second set of values of relaxation rates being $\gamma_{c} = 50\gamma$, $\gamma_{v} = 5 \gamma$. Each colors represent different values of $d$,  the tunneling parameter : $d=2$~nm (red lines), $d=4$~nm (blue lines), $d=6$~nm (green lines) and $d=10$~nm (orange lines). Gray dashed lines show the single quantum dot (SQD) calculations considering an energy gap $E_{12}$.}
	\label{fig2}
\end{figure}

We start our analysis by studying the behavior of the electronic current $j$ and power $P$ delivered by the QDM, varying the escape rates $\gamma_{c(v)}$ and thickness barrier $d$. In Fig.~\ref{fig2}, we show the current-voltage relation and the power as functions of voltages for several values of the separation $d$. The values of carriers escape rates $\gamma_{c}$ and $\gamma_{v}$, are set in terms of the relaxation rates, where $\gamma = \gamma_1 = \gamma_2$: the values $\gamma_{c} = 100\gamma$, $\gamma_{v} = 0.05\gamma$~\cite{Dorfman2011} were used in Figs.~\ref{fig2}(a) and \ref{fig2}(b), while values $\gamma_{c} = 50\gamma$, $\gamma_{v} = 5 \gamma$~\cite{Svidzinsky2011} were considered in the Figs.~\ref{fig2}(c) and \ref{fig2}(d). In both cases, the electron escape time is in the range of 10 ps to 20 ps, while the hole escape time is many orders of magnitude ($\sim 10^5$) slower~\cite{Fry2000,Brunkov2002}.

It is interesting to compare the behavior of the QDM with the case of a single quantum dot (SQD) of energy gap $E_{12}$, shown by black dashed lines in Fig.~\ref{fig2}. At first glance, the coherent tunneling leads to an enhancement on the photocurrent response. In both cases when $d$ decreases (see the red line for $d=2$ nm), and consequently the coherent tunneling rates increases, the short circuit current (maximum current) increases compared to the SQD maximum current. An interpretation of this results is that, when $d$ increases, the QDM behaves like a system of two decoupled QDs, with the inter-dot tunneling becoming an irrelevant process. This is illustrated in Figs.~\ref{fig2} (a) and (b) where, for $d = 10$ nm, the current and delivered power become closer to the dashed gray line of the SQD. The tunneling effects over $j$ and $P$ also depends on the values of $\gamma_{c}$ and $\gamma_{v}$ rates, as can be seen comparing Figs.~\ref{fig2} (c) and (d) with Figs.~\ref{fig2} (a) and (b) respectively.

\subsection{Effect of the relaxation rates}
It is interesting to analyze the combined action of the coherent tunneling and the relaxation rates, in order to quantify the increase of the photovoltaic response compared with the SQD case. We define the relative variation of the current as 
\begin{equation}
	\delta j = (j_{\mathrm{max}}^\mathrm{(QDM)} - j^\mathrm{(SQD)}) / j^\mathrm{(SQD)},
\end{equation}
where $j$ is chosen as the current for which the power reaches its maximum value $P_m$.  For $d=2$ nm and using the relaxation rates from Figs.~\ref{fig2} (a) and (c), we obtain the values $\delta j \sim 7\%$ and $\delta j \sim 31\%$, respectively. For the same interdot distance and using the relaxation rates from Figs.~\ref{fig2} (b) and (d), the QDM maximum power experiences a relative increase of $\Delta_{P_m} \sim 9\%$ and $\Delta_{P_m} \sim 32\%$, respectively. These results show the versatility of our physical system, in terms of the interplay between tunneling and the decay rates: for a fixed coupling of the QDM to the reservoirs, it is possible to obtain an increase in the photocurrent. Then, the power provided can be set controlling the size of the barrier.

The behavior of the physical system can be elucidated in terms of the main time scales. The radiative recombination rate, $\gamma$, defines the main time scale in the photovoltaic conversion process. When the recombination time $t_{\gamma} = h/\gamma$ is shorter than the relaxation times of electrons and holes, $t_{\gamma_{c}}$ and $t_{\gamma_v}$ to the contact states, the photovoltaic conversion can be inefficient since the radiative processes become dominant. If, on the contrary,  $t_{\gamma} \gg  t_{\gamma_c}, t_{\gamma_v}$, the transfer of carriers towards the contact states is the dominant effect, increasing the photocurrent.

\begin{figure}[htbp]
	\centering
	\includegraphics[scale=0.9]{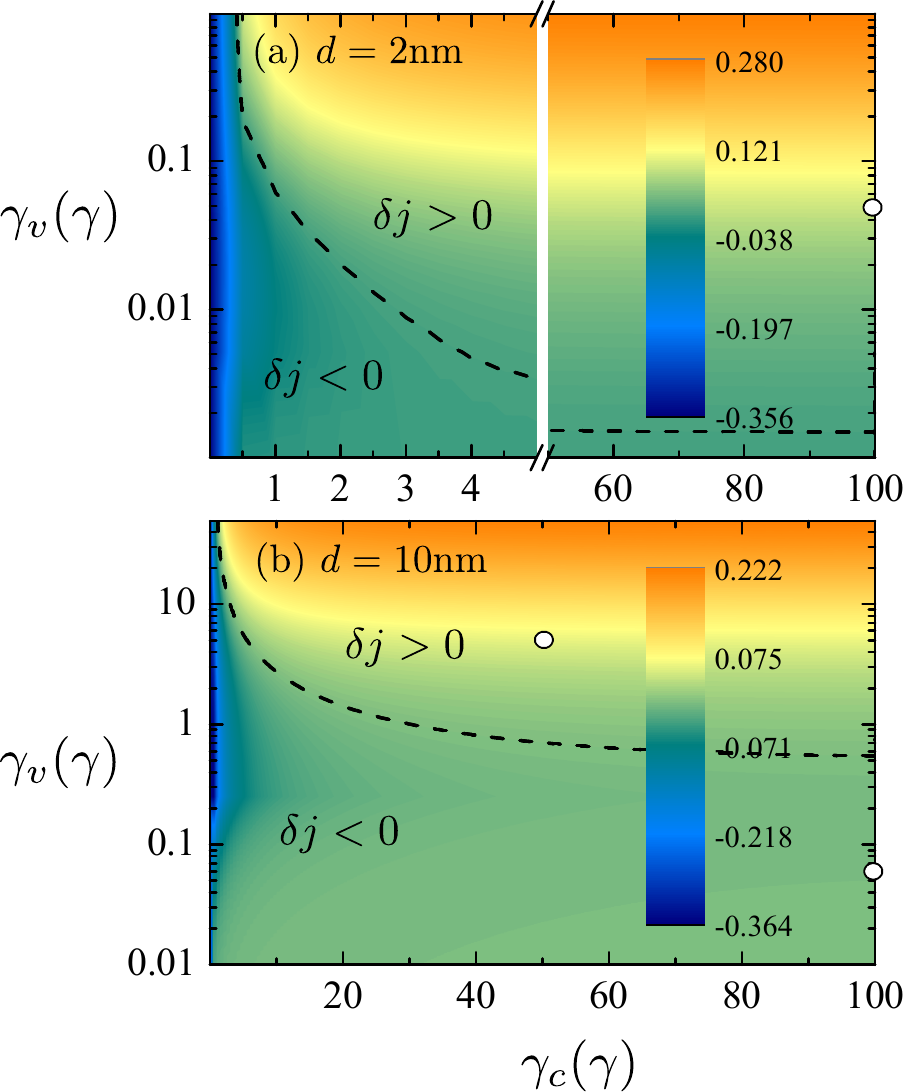}
	\caption{Relative variation of the current $\delta j$ for (a) $d=2$~nm and (b) $d=10$~nm, as a function of relaxation rates $\gamma_v$, $\gamma_{c}$. The dashed line represent $\delta j \simeq 0$. As  reference, the relaxation rates values used in Fig.~\ref{fig2} are indicated with white dots.}
	\label{fig3}
\end{figure}

To extend our analysis, it is important to find intervals of values for the relaxation parameters ($\gamma_{c}$ and $\gamma_v$), where the QDMs are more efficient in producing photocurrent than SQDs. In Fig.~\ref{fig3}, we show the relative variation of the current, $\delta j$, for two of the values of $d$ considered in Fig.~\ref{fig2}, as a function of the relaxation rates $\gamma_{c}$ and $\gamma_v$.
For $d = 2$ nm, and within values given by $\gamma_v > 0.001 \gamma$ and $\gamma_{c} \gg \gamma_v$, the current produced by the QDM will be appreciably higher than in SQD. Depending on the particular values assumed by  $\gamma_{c}$, $\gamma_v$, it is possible to obtain a current gain in the range of $\delta j$ between $10\%$ and $20\%$. The relative variation $\delta j$ does not increase indefinitely with increasing relaxation rates. If we keep the condition $\gamma_{c} \gg \gamma_v$ and with $\gamma_v \geq 5 \gamma$,  $\delta j$ reaches its maximum value of $\delta j \simeq 30\%$. Thus, the use of coherent tunneling to improve the generation of photocurrent and power is experimentally plausible, since in QDM it is generally expected that $\gamma_{c} \gg \gamma_v$. In the weak tunneling regime, the threshold of $\gamma_v$ for which $\delta j > 0$ increases by orders of magnitude compared to the $d = 2$~nm case. The region of $(\gamma_c, \gamma_v)$ parameters where the QDM current is greater than SQD current is shown in panel (b) for $d=10$~nm. In this regime, we obtain $\delta j>0$ when $\gamma_v > \gamma$ and $\gamma_{c} \gg \gamma_v$.

Our results show that it is favorable to use tunneled coupled structures instead of SQD to gain efficiency on the production of current. To qualitatively estimate the improvement in photovoltaic properties by using QDMs, it is necessary to solve equation (\ref{eq:scde}) and the corresponding density matrix equation of the SQD.  After some analytic calculations, we find an expression for the short circuit current to the first order of $(\gamma/\Gamma)$, which is written as 
\begin{equation}
	j^\mathrm{QDM} = \frac{(n_1 + n_2)(n_v + 1)}{(3n_v + 2)} + \mathcal{O}(1/T_e^2, 1/T_h^2).
\end{equation} 
The first term represents the current at the large tunneling limit, as tunneling coupling decreases the short circuit current also decreases. On the other hand, the short circuit current for a SQD to the first order of $(\gamma/\Gamma)$ is given by 
\begin{equation}
	j^\mathrm{SQD} = \frac{n_1(n_v +1)}{(2n_v + 1)}.    
\end{equation}
For large tunneling coupling, we compare the QDM and SQD shortcut current by means of the ratio $ j^\mathrm{QDM}/j^\mathrm{SQD} =  \frac{4n_v + 2}{3n_v + 2}$. This expression depends on the QDs operating temperature and establishes the upper limit of current gain produced by a QDM over the SQD. For identical QDs at room temperature, we obtain $ j^\mathrm{QDM}/j^\mathrm{SQD} \sim 4/3 = 1/\overline{3}$, which roughly represents a maximum gain of 30\% as suggested by our numerical results.

\subsection{Current, quantum coherences, and tunneling}
\begin{figure}[htbp]
	\centering
	\includegraphics[scale=0.6]{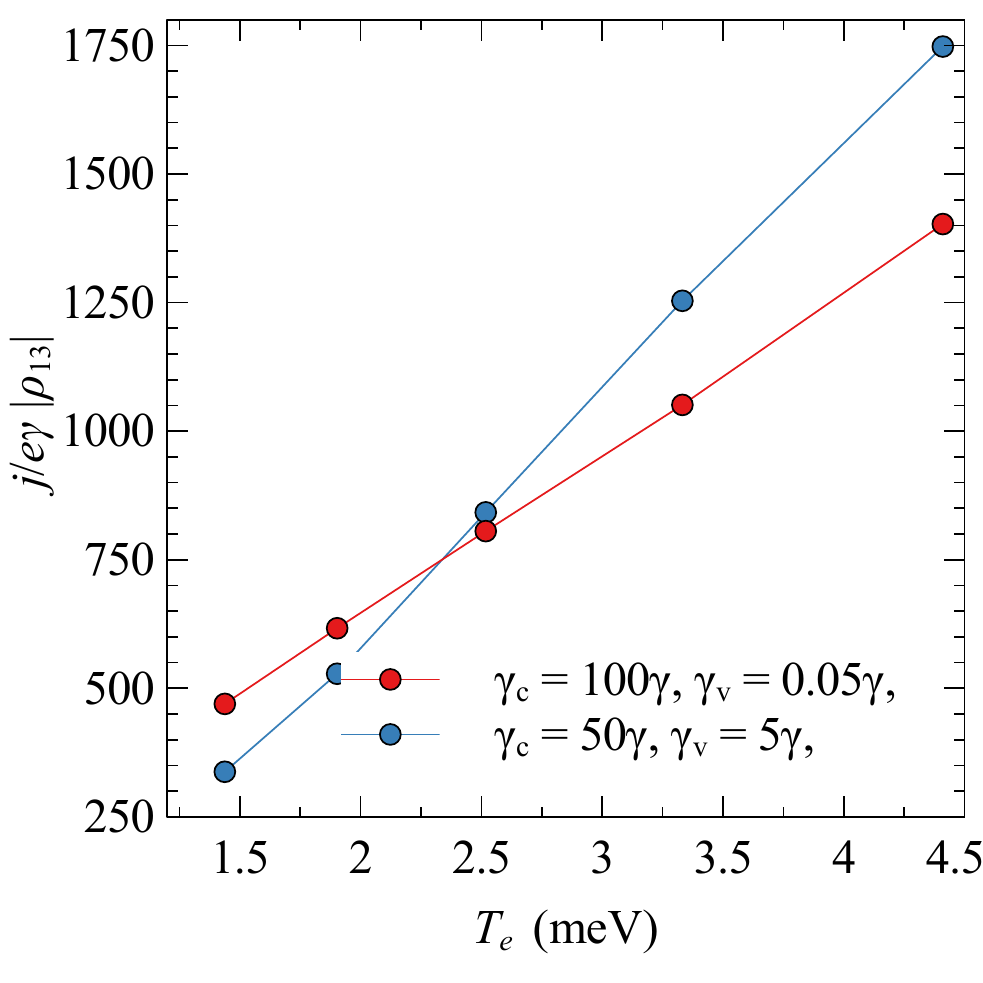}
	\caption{The ratio between the maximum current and coherence $\vert \rho_{13} \vert$ as a function of tunneling coupling $T_e$, showing the linear dependence of this quantity with $T_e$ in the two considered cases: $\gamma_{c} = 100\gamma$, $\gamma_{v} = 0.05\gamma$ (red line and dots), and $\gamma_{c} = 50\gamma$, $\gamma_{v} = 5\gamma$ (blue line and dots).}
	\label{fig4}
\end{figure}

It is important to examine the effect that the quantum coherences $\rho_{ij}$ have on the generation of the charge current in QDM's. For this purpose, we explore an analitical solution consider the system of coupled differential equations in Appendix~\ref{app:coupleddeq} in the limit where $\gamma \ll \Gamma$. This allows to obtain an approximate solution for $\rho_{13}$ and find a relation with $j^\mathrm{QDM}$ which is written in a simplified form as 
\begin{equation}
	j^\mathrm{QDM} \propto \frac{f(\gamma_i)}{\sqrt{g(\gamma_i)^2 + \Delta_e^2}}T_e \vert \rho_{13}\vert,    
\end{equation}
where $f(\gamma_i)$ and $g(\gamma_i)$ are constants which is fixed by the value of the damping $\gamma_{c(v)}$ and the occupations of the levels. This expression shows that the current is directly proportional to the magnitude of the coherence, with the $T_e$ prefactor allowing the current to reach its maximum value when $T_e$ increases. We numerically verify this linear dependence in Fig.~\ref{fig4}, where we plot the ratio between the maximum current and coherence, $\frac{j}{e\gamma \vert \rho_{13} \vert}$, which is proportional to tunneling coupling $T_e$. Notice that the slope of the illustrated case depends on the values of $\gamma_{c(v)}$, with the higher slope (blue line and dots) corresponding to the higher value of the relaxation rate $\gamma_v$.

\begin{figure}[htbp]
	\centering
	\includegraphics[scale=0.5]{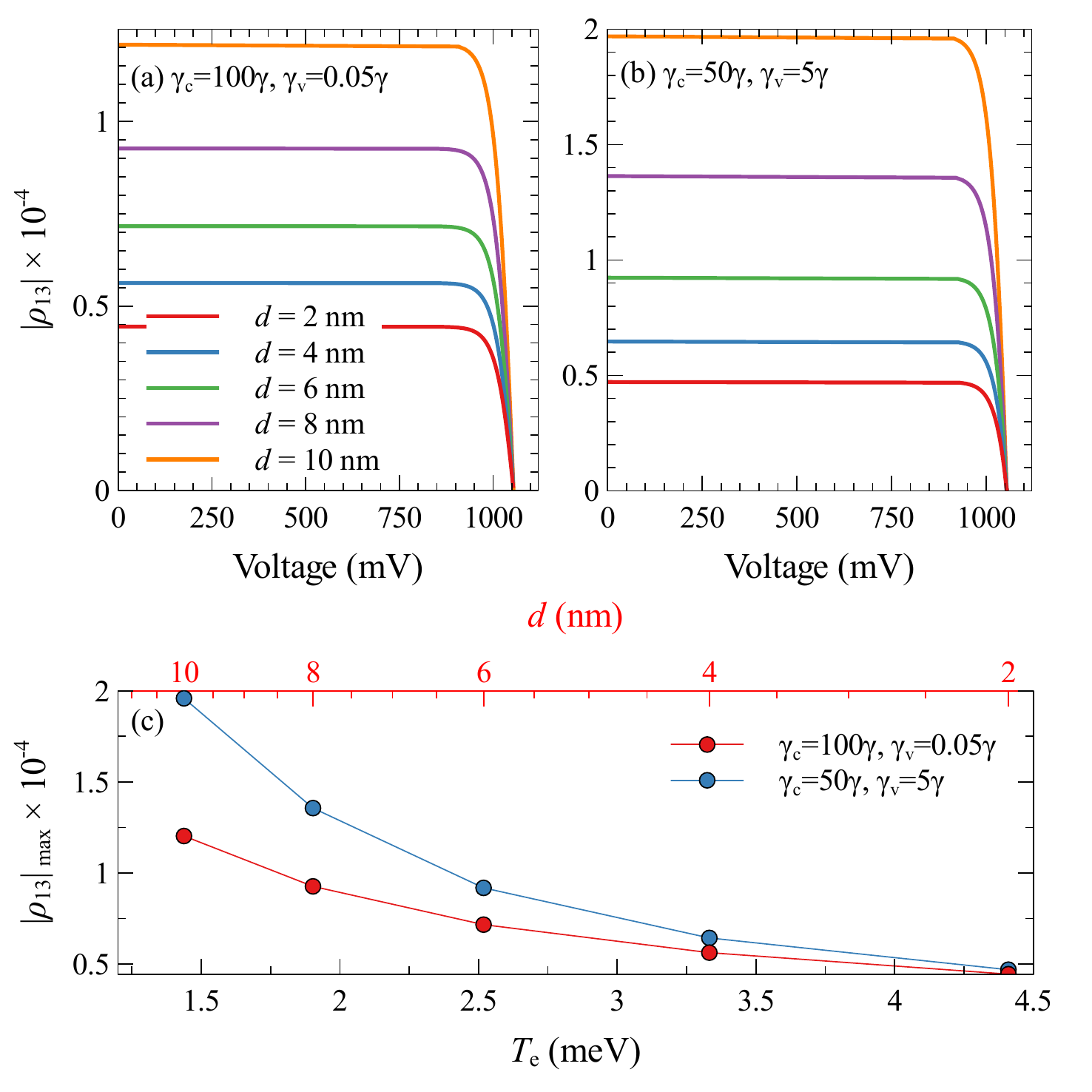}
	\caption{The coherence term $\vert \rho_{13} \vert$ as a function of the induced voltage considering (a) $\gamma_{c} = 100\gamma$, $\gamma_{v} = 0.05\gamma$, and (b) $\gamma_{c} = 50\gamma$, $\gamma_{v} = 5\gamma$, for several values of interdot separations: $d=2$~nm (solid red line), $d=4$~nm (solid blue line), $d=6$~nm (green red line), $d=8$~nm (solid brown line), and $d=10$~nm (solid orange line). Panel (c): maximum value of the coherence as a function of interdot separation $d$ and its corresponding tunneling coupling considering $\gamma_{c} = 100\gamma$, $\gamma_{v} = 0.05\gamma$ (red line and dots) and $\gamma_{c} = 50\gamma$, $\gamma_{v} = 5\gamma$ (blue line dots).}
	\label{fig5}
\end{figure}

To investigate the dependence of the coherence with $T_e$, we numerically calculate $\vert \rho_{13}\vert$ for different values of the interdot separation and their corresponding tunneling couplings. Our results are summarized in Fig. \ref{fig5}, where two sets of relaxation parameters $(\gamma_c, \gamma_v)$ were considered. As expected, when comparing the panels (a) and (b), the higher the relaxation rates, the lower the coherence. From panel (c) we can see how, for the range of values of $T_e$ considered, the coherence is a decreasing function of tunneling.

This counterintuitive result can be explained with the aid of a two-level system (TLS) model~\footnote{In a TLS model with phenomenological population $\gamma_0$ and coherence $\gamma^\prime$ damping rates, the steady-state population and coherence are given by $\rho_{ee} = \frac{W^2}{2\gamma_0 \gamma^\prime} \frac{1}{1 + \frac{\delta^2}{{\gamma^\prime}^2} + \frac{W^2}{\gamma_0 \gamma^\prime} }$, $\rho_{eg} = -i \frac{W}{2 \gamma^\prime} \frac{1 + \frac{i\delta}{\gamma^\prime}}{1 + \frac{\delta^2 }{{\gamma^\prime}^2} + \frac{W^2}{\gamma_0 \gamma^\prime}}$, respectively.}. Let us consider $\gamma_0$ and $\gamma^\prime$ as the population and coherence damping rates respectively, of a TLS driven by a $W$ interaction. The TLS population and coherence are related by 
\begin{equation}
	\rho_{ee}= \frac{\gamma^\prime}{\gamma_0} \frac{W}{\sqrt{{\gamma^\prime}^2 + \delta^2}}\vert \rho_{eg}\vert,     
\end{equation}
where $\delta$ is the TLS transition energy. This relation is functionally consistent with our results and captures the essential aspects of the coherence effects on the delivered current. In the TLS stationary regime, it can be shown that when $W$ increases both quantities,  the population $\rho_{ee}$ and the coherence $\rho_{eg}$, also increase. As $W$ continues to increase, the population tends to saturate and when 
\begin{equation}
	W> W^\prime = \sqrt{\gamma_0/\gamma^\prime} \sqrt{\delta^2 + {\gamma^\prime}^2}    
\end{equation}
the coherence progressively decreases. For sufficiently large coupling values, $W \gg W^\prime $, the coherence asymptotically tends to zero and the population is independent of the TLS coupling $W$.

The discussion presented here let us conclude that for a QDM system, the states $\vert 1 \rangle$ and $\vert 3 \rangle$ interact through an effective coupling, which involves the tunneling coupling $T_e$. This effective coupling is large enough to drive the population near the saturation threshold. Under these conditions, the current increases slowly with increasing $T_e$ and in turn, the coherence decreases progressively, as can be seen in Fig.~\ref{fig5}(c). Therefore, for the experimentally realistic QDMs studied here, the increase in current is not directly related to an improvement of the coherent processes. For weakly coupled quantum dots in a QDM or large transition energies, we can expect a direct connection between the delivered current and coherent mechanisms.

On the other hand, since $T_e > T_h$ the valence states $\vert 2 \rangle$ and $\vert 4 \rangle$ are weakly coupled by tunneling compared to states $\vert 1 \rangle $ and $\vert 3 \rangle$. For the experimental values of $T_h$ considered here, and with $\gamma_v = 0.005 \gamma$  we confirm numerically that the coherence $\vert \rho_{24}\vert $ is an increasing function of $T_h$ (not presented here). Furthermore, since the relaxation rate for state $\vert 6 \rangle$, $\gamma_v$, is very small as compared to the rate $\gamma_c$, the coherence $\vert \rho_{24}\vert $ is about two orders of magnitude higher than $\vert \rho_{13} \vert $. Consequently, the population transfer from state $\vert 2 \rangle$ to state $\vert 6 \rangle $ is efficiently driven by coherence.

Finally, we briefly address the effect of phonon-assisted tunneling on the photovoltaic properties of the QDM, a process has been widely studied in the literature~\cite{Gawarecki10,Muller12,Nakaoka06,Stace05,Gauger08} in the context of semiconductor nanoestructures. In order to taking into account the incoherent process, we include an additional term on Eq.~(\ref{eq:Lindblad}), with a similar structure of Eq.~(\ref{eq:liovillian}), considering different values for the phonon-assisted tunneling rates $\gamma_{13}$ (conduction states), and $\gamma_{24}$ (valence states). To evaluate the effect, we perform numerical calculations for the power generated, $P/\gamma$, as a function of voltage considering three different values of rates being $\gamma_{13} = \gamma_{24} = 0.001\gamma,0.01\gamma,$ and $0.1\gamma$. Our calculations show that the general behavior of the photovoltaic properties  does not differ from the one shown in Fig.~\ref{fig2}(b). Still, if compare with the case presented in Fig.~\ref{fig2}(b), $P/\gamma$ shows a gain of $7.7\%$ for $d = 2$ nm and $14.7\%$ for $d = 10$ nm for $\gamma_{13} = \gamma_{24}=0.001\gamma$. For same the physical parameters used in Fig.~\ref{fig2}(d), $P/\gamma$ remains unchanged for $d = 2$ nm, but shows a gain of $10.4\%$ for $d = 10$ nm. When the rates $\gamma_{13}$ and $\gamma_{24}$ increase, the values of $P/\gamma$ continue to increase, with the maximum power becoming constant, regardless of the value of the barrier $d$.

\section{Band alignment effects and efficiency}
\label{sec:bands}
In this section, we discuss the effect of the alignments (detunings) of electrodes and electron and holes energy levels. Once today's growth techniques allow accurate engineering of the QDM band alignments, it is important to address the effect of this feature on the photovoltaic properties of QDM. We explore four energetic configurations, represented schematically in Fig.~\ref{fig1}(b), and defined in Table~\ref{table1}, where the values of conduction and valence QDM band offsets was kept fixed at $\Delta_c = \Delta_v = 2$ meV. We include in the analysis the band alignment identified with the subscript ``$0$", which refers to the set of parameters used in the Section~\ref{sec:cohtunnel}. Here the subindex $1$ indicates resonance conditions associated with the conduction reservoir, and electronic states $|1\rangle$ and $|3\rangle$, while subindex $2$ refers to similar choices concerning the valence reservoir, and hole states $|2\rangle$ and $|4\rangle$.

\begin{table}[t]
	\centering\begin{tabular}{|l|l|l|l|l|}
		\hline
		& $A_1$ & $A_2$ & $B_1$ & $B_2$  \\
		\hline \hline
		$\Delta_e$ &  $ 0 $ & $\Delta_{h}^0 + \Delta_e^0$ & $-\Delta_c$ & $ \Delta_e^0 + \Delta_h^0 - \Delta_v$ \\
		\hline
		$\Delta_h$ &  $\Delta_h^0 + \Delta_e^0$ & $0 $ & $\Delta_h^0 + \Delta_e^0 + \Delta_c$ & $ \Delta_v$ \\
		\hline
	\end{tabular}
	\caption{The four QDM conduction, and valence band alignment parameters, as shown schematically in Fig.~\ref{fig1}(b). For numerical simulations, we consider the values $\Delta_e^0=\Delta_h^0= 3$ meV, and $\Delta_c=\Delta_v= 2$ meV.}
	\label{table1}
\end{table}

\begin{figure}[htbp]
	\centering
	\includegraphics[scale=0.5]{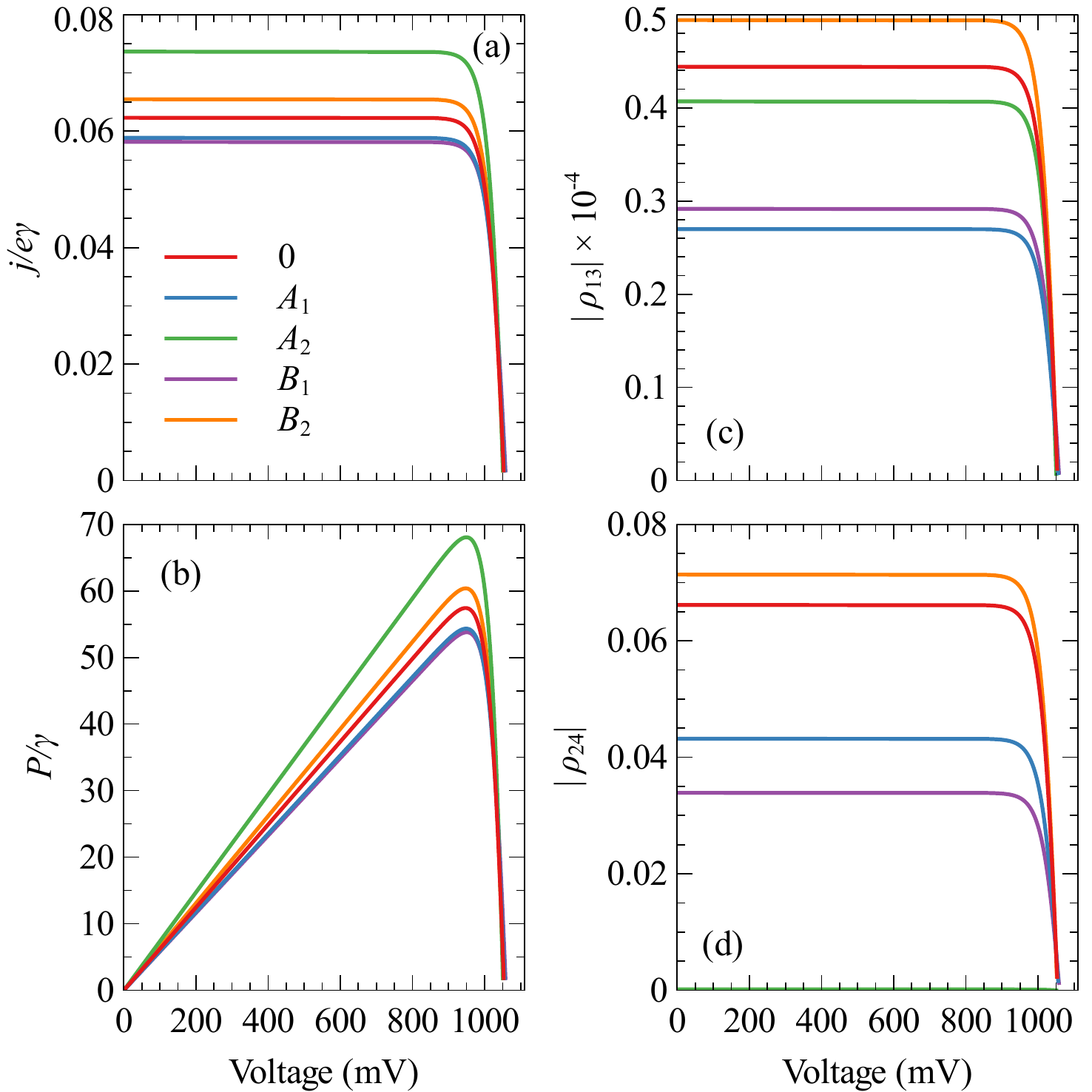}
	\caption{(Color online) (a) QDM characteristic current-voltage relation and, (b) the power generated as functions of induced voltage considering the inderdot separation as $d=2$ nm, and relaxation rates $\gamma_{c} = 100 \gamma$ and $\gamma_{v} = 0.05 \gamma$. Panels (c) and (d) show the behavior of coherences terms $\vert \rho_{13}\vert $ and $\vert \rho_{24}\vert$ also as a functions of induced voltage. Each color represents one band alignment configurations from Table~\ref{table1}: $A_1$ (blue line), $A_2$ (green line), $B_1$ (purple line), and $B_2$ (orange line) compared with the reference case ``$0$" (red line) from Figs.~\ref{fig2}, and~\ref{fig3}.}
	\label{fig6}
\end{figure}

The behaviors of current, power and the coherences $\vert \rho_{13}\vert $ and $\vert \rho_{24}\vert$ for the alignments in Table~\ref{table1} are shown in Fig.~\ref{fig6}, considering $d=2$ nm. We note that configuration $A_2$ delivers the highest current, Fig.~\ref{fig6}(a), and the highest power, Fig.~\ref{fig6}(b), while the configuration $B_2$ shows the highest values of the coherence terms $\vert \rho_{13}\vert$, Fig.~\ref{fig6}(c), and $\vert \rho_{24}\vert$, Fig.~\ref{fig6}(d). Notice that the configuration $A_2$ is associated with a resonance condition for the states $|2\rangle$ and $|4\rangle$, showing a very low value of the coherence term $\vert \rho_{24}\vert$ ($\approx 0.2\times 10^{-2}$), shown by a green line slightly above the  $x$-axis in Fig.~\ref{fig6}(d).

\begin{figure}[htbp]
	\centering
	\includegraphics[scale=0.5]{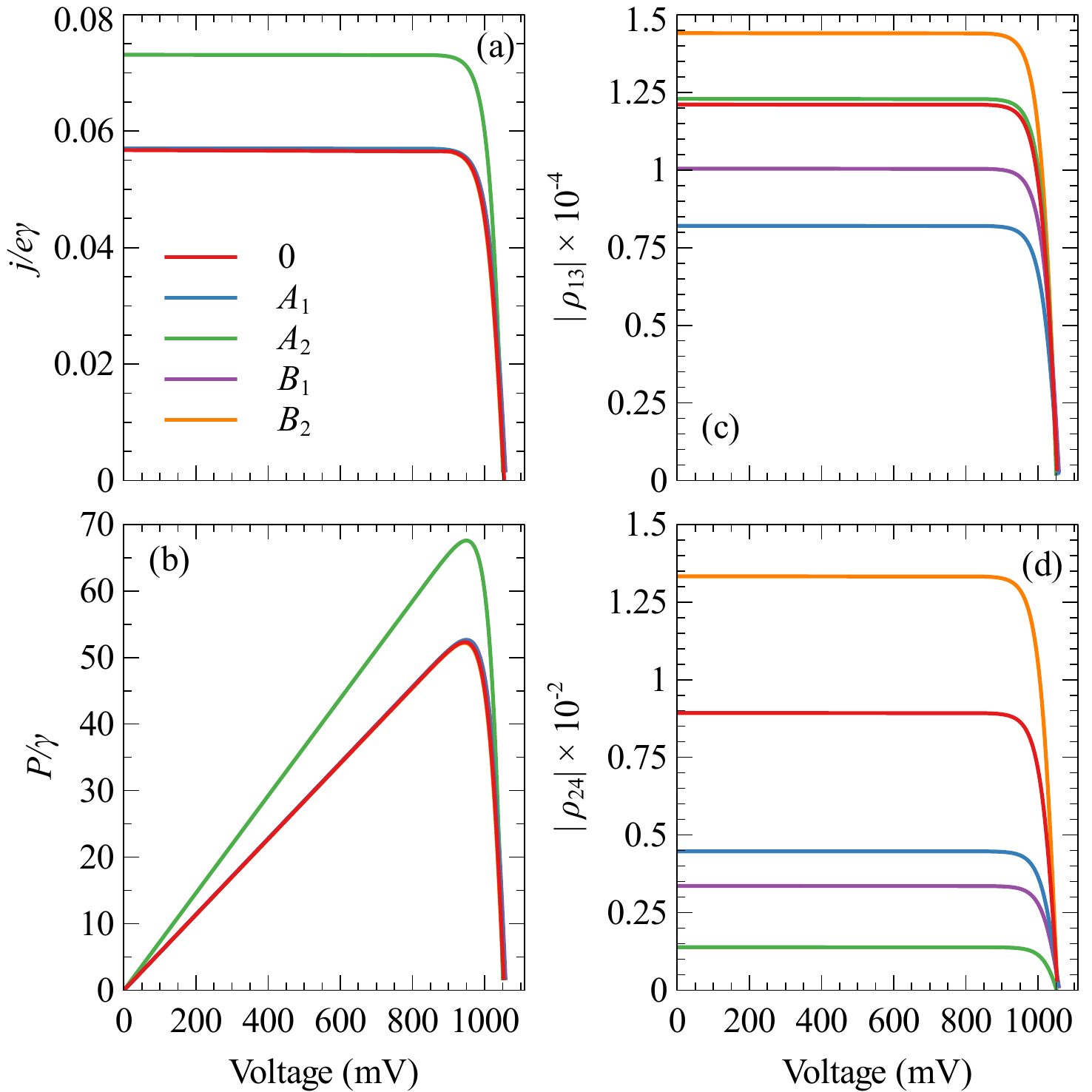}
	\caption{(Color online) (a) QDM characteristic current-voltage relation and, (b) the power generated as functions of induced voltage considering the inderdot separation as $d=10$ nm, and relaxation rates $\gamma_{c} = 100 \gamma$ and $\gamma_{v} = 0.05 \gamma$. Panels (c) and (d) show the behavior of coherences $\vert \rho_{13}\vert $ and $\vert \rho_{24}\vert$ also as a functions of induced voltage. Each color represents one band alignment configurations from Table~\ref{table1}: $A_1$ (blue line), $A_2$ (green line), $B_1$ (purple line), and $B_2$ (orange line) compared with the reference case ``$0$" (red line)from Figs.~\ref{fig2}, and~\ref{fig3}.}
	\label{fig7}
\end{figure}

In Fig.~\ref{fig7} we show the behavior of current, power and the coherence terms for $d=10$ nm, in order to explore the role of the interdot separation. The results show that the configuration $A_2$ still provides the higher values of current and power, panels (a) and (b), with similar values to those found for $d=2$ nm. However, the alignments apart from $A_2$ fall in the same curve of current and power associated with $A_1$ and $B_1$ in Fig.~\ref{fig6}. Comparing the results, one can say that for configurations $A_1$, $B_1$, and $A_2$, the current is practically independent of the interdot separation. These configurations are characterized by the presence of resonances either in the conduction or in the valence states. Regardless of the values of the relaxation rates, this particular choice of resonances causes the populations to quickly reach the saturation threshold even for small values of tunneling. Thus, the current no longer would depend on the behavior of the coherence.

About the behavior of the coherence terms, we see that $\vert \rho_{13} \vert $ increases with $d$ for all the band alignments, if we compare the panels (c) from both figures, Fig.~\ref{fig6} and Fig.~\ref{fig7}. This behavior is explained in terms on the effect of the tunneling coupling, which leads the populations close to saturation where the coherence decreases when $T_e$ increases. On the other hand, the term $\vert \rho_{24} \vert $ decreases rapidly with increasing distance $d$, panels (d) of Fig.~\ref{fig6} and Fig.~\ref{fig7}. Here, $T_h$ coupling is sufficiently weak so that the population does not reach the saturation limit and therefore the coherence gradually decreases as $T_h$ increases. Finally, we investigated the efficiency of the QDM for all the band alignment configurations using Eq.~(\ref{eq:efficiency}). For all the alignments considered, $P_S$ is always higher than the maximum power delivered and the efficiency is limited by the Carnot efficiency so $\eta=P_m/P_S \leq 1 + T_c/T_S$.

\begin{figure}[htbp]
	\centering
	\includegraphics[scale=0.5]{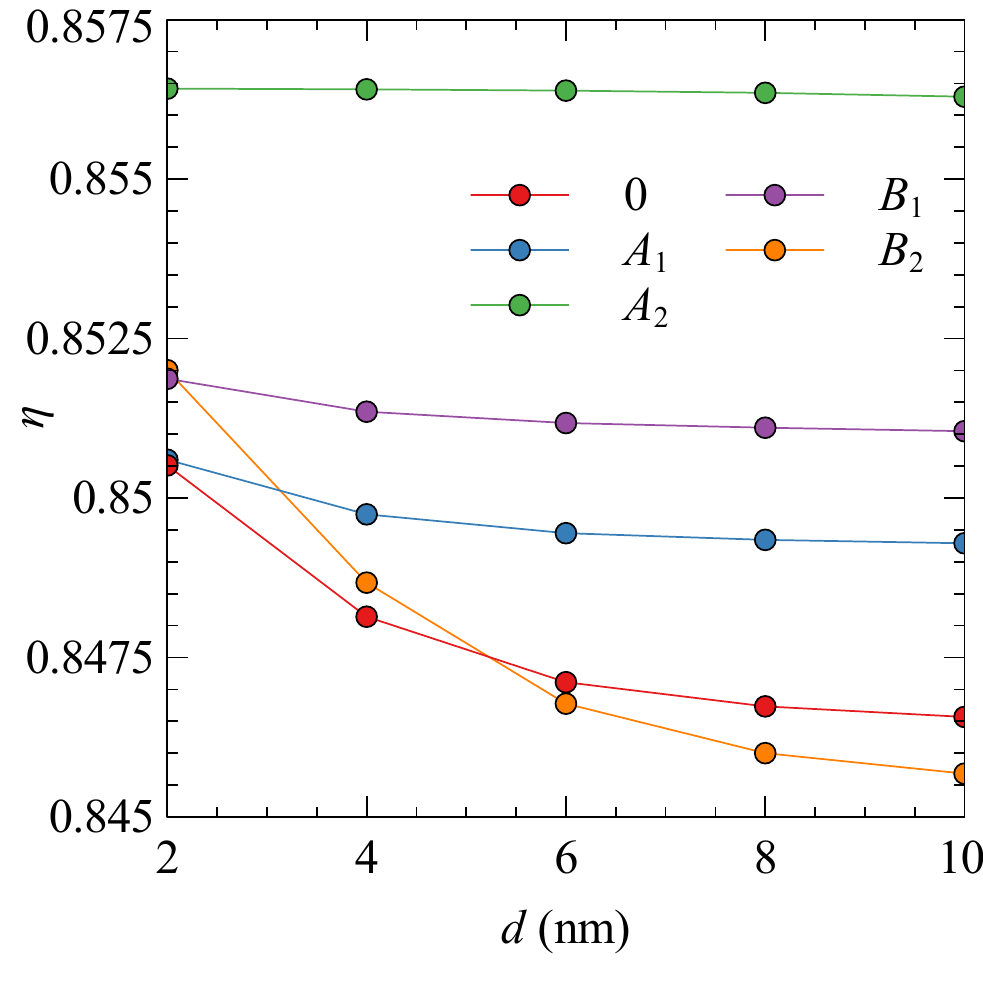}
	\caption{The efficiency, $\eta$, of the QDM calculated as a function of the tunneling barrier width $d$. We consider $\gamma_{c} = 100 \gamma$ and $\gamma_{v} = 0.05 \gamma$. The color scheme is the same used in Fig.~\ref{fig6}, but now using a solid line and dot plot style.}
	\label{fig8}
\end{figure}

Our results are summarized in Fig.~\ref{fig8}. We notice that the QDM in the $A_2$ configuration delivers the largest power, which is in agreement with Figs.~\ref{fig6}(b) and \ref{fig7} (b). We also note that the efficiencies for configurations $A_1$, $B_1$ and $A_2$ are weakly dependent on the interdot separation $d$. This is because for these configurations, the coherent effects on photovoltaic processes are weak, as discussed before. In contrast, for configurations $0$ and $B_2$, where coherent effects are important, the efficiency shows a clear dependence on $d$ and consequently on the tunneling coupling.

\section{Summary}
\label{sec:summary}
We theoretically investigate the effects of coherent tunneling on the photovoltaic properties of a quantum dot molecule coupled with a hot reservoir, being the solar radiation, and a cold reservoir, being the thermal phonons of the semiconductor where the device is embedded. The main goal was to establish the relationship between the generation of coherence due to tunneling and its impact on the photocurrent and power generated by the photocell. In our analysis, we use realistic values for all the physical parameters, including the tunneling rates, taken from experimental reports.

Our results show that the effects of coherent tunneling improve the current generation and power delivered in a range of 8\% and 30\% if compared with a single quantum dot.  To assist us in our analysis, we use a two-level system model, which allowed us to establish that the amount of coherence, in the steady-state, is not a monotonically increasing function of the tunneling coupling. We also show that the photovoltaic response depends significantly on the thermal phonon coupling rates, which connect the QDM states with the contact states. We check that the phonon-assisted (incoherent) tunneling does not produce important changes in the photovoltaic behavior. We finally analyze the effect of manipulation of the detunings on the generation of current, power, and the behavior of the efficiency, obtaining an ideal condition of these physical parameters for a potential application of the QDM in the conversion of solar light.

\section{Acknowledgments}
This work was supported by CAPES, the Conselho Nacional de Desenvolvimento Científico e Tecnológico (CNPq), grant 422350/2021-4, the Fundação de Apoio Universitário (edital 02/2018), and the Brazilian National Institute of Science and Technology of Quantum Information (INCT-IQ), grant 465469/2014-0/CNPq.
\appendix
\section{The master equations for the model of a quantum molecule as a photocell}
\label{app:coupleddeq}
The Eq.~(\ref{eq:Lindblad}) results on a system of coupled differential equations, which depends on the physical parameters of the system. The equation are written as:
\begin{eqnarray}
\dot{\rho}_{11}&=&-\frac{i}{\hbar}T_e(\rho_{31}-\rho_{13})-\gamma_1[(n_1+1)\rho_{11}-n_1\rho_{22}],\nonumber\\
\dot{\rho}_{22}&=&-\frac{i}{\hbar}T_h(\rho_{42}-\rho_{24})-\gamma_1[-(n_1+1)\rho_{11}+n_1\rho_{22}]\nonumber\\
&&-\gamma_v[-(n_v+1)\rho_{vv}+n_v\rho_{22}],\nonumber\\
\dot{\rho}_{33}&=&-\frac{i}{\hbar}T_e(-\rho_{31}+\rho_{13})-\gamma_2[(n_2+1)\rho_{33}-n_2\rho_{44}]\nonumber\\
&&-\gamma_c[(n_c+1)\rho_{33}-n_c\rho_{cc}],\nonumber\\
\dot{\rho}_{44}&=&-\frac{i}{\hbar}T_h(-\rho_{42}+\rho_{24})-\gamma_2[-(n_2+1)\rho_{33}+n_2\rho_{44}],\nonumber
\end{eqnarray}
\begin{eqnarray}
\dot{\rho}_{13}&=&-i(\omega_1 - \omega_3)\rho_{13} -\frac{i}{\hbar}T_e(\rho_{33}-\rho_{11})\nonumber\\
&&-\frac{1}{2}[\gamma_1(n_1+1)+\gamma_2(n_2+1)+\gamma_c(n_c+1)]\rho_{13},\nonumber\\
\dot{\rho}_{31}&=&-i(\omega_3 - \omega_1)\rho_{31} -\frac{i}{\hbar}T_e(\rho_{11}-\rho_{33})\nonumber\\ &&-\frac{1}{2}[\gamma_1(n_1+1)+\gamma_2(n_2+1)+\gamma_c(n_c+1)]\rho_{31},\nonumber\\
\dot{\rho}_{24}&=&-i(\omega_2-\omega_4)\rho_{24} -\frac{i}{\hbar}T_h(\rho_{44}-\rho_{22})\nonumber\\
&&-\frac{1}{2}[\gamma_1 n_1+ \gamma_2 n_2+\gamma_v n_v]\rho_{24},\nonumber\\
\dot{\rho}_{42}&=&-i(\omega_4-\omega_2)\rho_{42} -\frac{i}{\hbar}T_h(\rho_{22}-\rho_{44})\nonumber\\
&&-\frac{1}{2}[\gamma_1 n_1+ \gamma_2 n_2+\gamma_v n_v]\rho_{42},\nonumber\\
\dot{\rho}_{cc}&=&-\gamma_c[-(n_c+1)\rho_{33}+n_c\rho_{cc}]-\Gamma\rho_{cc},\nonumber\\
\dot{\rho}_{vv}&=&-\gamma_v[(n_v+1)\rho_{vv}-n_v\rho_{22}]+\Gamma\rho_{cc}.\label{eq:scde}
\end{eqnarray}

%\bibliography{solarqdnew_2rev}

\begin{thebibliography}{42}%
	\makeatletter
	\providecommand \@ifxundefined [1]{%
		\@ifx{#1\undefined}
	}%
	\providecommand \@ifnum [1]{%
		\ifnum #1\expandafter \@firstoftwo
		\else \expandafter \@secondoftwo
		\fi
	}%
	\providecommand \@ifx [1]{%
		\ifx #1\expandafter \@firstoftwo
		\else \expandafter \@secondoftwo
		\fi
	}%
	\providecommand \natexlab [1]{#1}%
	\providecommand \enquote  [1]{``#1''}%
	\providecommand \bibnamefont  [1]{#1}%
	\providecommand \bibfnamefont [1]{#1}%
	\providecommand \citenamefont [1]{#1}%
	\providecommand \href@noop [0]{\@secondoftwo}%
	\providecommand \href [0]{\begingroup \@sanitize@url \@href}%
	\providecommand \@href[1]{\@@startlink{#1}\@@href}%
	\providecommand \@@href[1]{\endgroup#1\@@endlink}%
	\providecommand \@sanitize@url [0]{\catcode `\\12\catcode `\$12\catcode
		`\&12\catcode `\#12\catcode `\^12\catcode `\_12\catcode `\%12\relax}%
	\providecommand \@@startlink[1]{}%
	\providecommand \@@endlink[0]{}%
	\providecommand \url  [0]{\begingroup\@sanitize@url \@url }%
	\providecommand \@url [1]{\endgroup\@href {#1}{\urlprefix }}%
	\providecommand \urlprefix  [0]{URL }%
	\providecommand \Eprint [0]{\href }%
	\providecommand \doibase [0]{https://doi.org/}%
	\providecommand \selectlanguage [0]{\@gobble}%
	\providecommand \bibinfo  [0]{\@secondoftwo}%
	\providecommand \bibfield  [0]{\@secondoftwo}%
	\providecommand \translation [1]{[#1]}%
	\providecommand \BibitemOpen [0]{}%
	\providecommand \bibitemStop [0]{}%
	\providecommand \bibitemNoStop [0]{.\EOS\space}%
	\providecommand \EOS [0]{\spacefactor3000\relax}%
	\providecommand \BibitemShut  [1]{\csname bibitem#1\endcsname}%
	\let\auto@bib@innerbib\@empty
	%</preamble>
	\bibitem [{\citenamefont {Shockley}\ and\ \citenamefont
		{Queisser}(1961)}]{Shockley1961}%
	\BibitemOpen
	\bibfield  {author} {\bibinfo {author} {\bibfnamefont {W.}~\bibnamefont
			{Shockley}}\ and\ \bibinfo {author} {\bibfnamefont {H.~J.}\ \bibnamefont
			{Queisser}},\ }\bibfield  {title} {\bibinfo {title} {Detailed balance limit
			of efficiency of p-n junction solar cells},\ }\href
	{https://doi.org/10.1063/1.1736034} {\bibfield  {journal} {\bibinfo
			{journal} {Journal of Applied Physics}\ }\textbf {\bibinfo {volume} {32}},\
		\bibinfo {pages} {510} (\bibinfo {year} {1961})}\BibitemShut {NoStop}%
	\bibitem [{\citenamefont {Zhao}\ and\ \citenamefont {Chen}(2019)}]{Zhao2019}%
	\BibitemOpen
	\bibfield  {author} {\bibinfo {author} {\bibfnamefont {S.-C.}\ \bibnamefont
			{Zhao}}\ and\ \bibinfo {author} {\bibfnamefont {J.-Y.}\ \bibnamefont
			{Chen}},\ }\bibfield  {title} {\bibinfo {title} {Enhanced quantum yields and
			efficiency in a quantum dot photocell modeled by a multi-level system},\
	}\href@noop {} {\bibfield  {journal} {\bibinfo  {journal} {New J. Phys.}\
		}\textbf {\bibinfo {volume} {21}},\ \bibinfo {pages} {103015} (\bibinfo
		{year} {2019})}\BibitemShut {NoStop}%
	\bibitem [{\citenamefont {Dorfman}\ \emph {et~al.}(2018)\citenamefont
		{Dorfman}, \citenamefont {Xu},\ and\ \citenamefont {Cao}}]{Dorfman2018}%
	\BibitemOpen
	\bibfield  {author} {\bibinfo {author} {\bibfnamefont {K.~E.}\ \bibnamefont
			{Dorfman}}, \bibinfo {author} {\bibfnamefont {D.}~\bibnamefont {Xu}},\ and\
		\bibinfo {author} {\bibfnamefont {J.}~\bibnamefont {Cao}},\ }\bibfield
	{title} {\bibinfo {title} {{Efficiency at Maximum Power of a Laser Quantum
				Heat Engine Enhanced by Noise-Induced Coherence}},\ }\href
	{https://doi.org/10.1103/PhysRevE.97.042120} {\bibfield  {journal} {\bibinfo
			{journal} {Phys. Rev. E}\ }\textbf {\bibinfo {volume} {97}},\ \bibinfo
		{pages} {042120} (\bibinfo {year} {2018})}\BibitemShut {NoStop}%
	\bibitem [{\citenamefont {Creatore}\ \emph {et~al.}(2013)\citenamefont
		{Creatore}, \citenamefont {Parker}, \citenamefont {Emmott},\ and\
		\citenamefont {Chin}}]{Creatore2013}%
	\BibitemOpen
	\bibfield  {author} {\bibinfo {author} {\bibfnamefont {C.}~\bibnamefont
			{Creatore}}, \bibinfo {author} {\bibfnamefont {M.~A.}\ \bibnamefont
			{Parker}}, \bibinfo {author} {\bibfnamefont {S.}~\bibnamefont {Emmott}},\
		and\ \bibinfo {author} {\bibfnamefont {A.~W.}\ \bibnamefont {Chin}},\
	}\bibfield  {title} {\bibinfo {title} {{Efficient Biologically Inspired
				Photocell Enhanced by Delocalized Quantum States}},\ }\href
	{https://doi.org/10.1103/PhysRevLett.111.253601} {\bibfield  {journal}
		{\bibinfo  {journal} {Phys. Rev. Lett.}\ }\textbf {\bibinfo {volume} {111}},\
		\bibinfo {pages} {253601} (\bibinfo {year} {2013})}\BibitemShut {NoStop}%
	\bibitem [{\citenamefont {Scully}(2010)}]{Scully2010}%
	\BibitemOpen
	\bibfield  {author} {\bibinfo {author} {\bibfnamefont {M.~O.}\ \bibnamefont
			{Scully}},\ }\bibfield  {title} {\bibinfo {title} {Quantum photocell: Using
			quantum coherence to reduce radiative recombination and increase
			efficiency},\ }\href {https://doi.org/10.1103/PhysRevLett.104.207701}
	{\bibfield  {journal} {\bibinfo  {journal} {Phys. Rev. Lett.}\ }\textbf
		{\bibinfo {volume} {104}},\ \bibinfo {pages} {207701} (\bibinfo {year}
		{2010})}\BibitemShut {NoStop}%
	\bibitem [{\citenamefont {Dorfman}\ \emph {et~al.}(2011)\citenamefont
		{Dorfman}, \citenamefont {Kim},\ and\ \citenamefont
		{Svidzinsky}}]{Dorfman2011}%
	\BibitemOpen
	\bibfield  {author} {\bibinfo {author} {\bibfnamefont {K.~E.}\ \bibnamefont
			{Dorfman}}, \bibinfo {author} {\bibfnamefont {M.~B.}\ \bibnamefont {Kim}},\
		and\ \bibinfo {author} {\bibfnamefont {A.~A.}\ \bibnamefont {Svidzinsky}},\
	}\bibfield  {title} {\bibinfo {title} {{Increasing photocell power by quantum
				coherence induced by external source}},\ }\href
	{https://doi.org/10.1103/PhysRevA.84.053829} {\bibfield  {journal} {\bibinfo
			{journal} {Phys. Rev. A}\ }\textbf {\bibinfo {volume} {84}},\ \bibinfo
		{pages} {53829} (\bibinfo {year} {2011})}\BibitemShut {NoStop}%
	\bibitem [{\citenamefont {Svidzinsky}\ \emph {et~al.}(2011)\citenamefont
		{Svidzinsky}, \citenamefont {Dorfman},\ and\ \citenamefont
		{Scully}}]{Svidzinsky2011}%
	\BibitemOpen
	\bibfield  {author} {\bibinfo {author} {\bibfnamefont {A.~A.}\ \bibnamefont
			{Svidzinsky}}, \bibinfo {author} {\bibfnamefont {K.~E.}\ \bibnamefont
			{Dorfman}},\ and\ \bibinfo {author} {\bibfnamefont {M.~O.}\ \bibnamefont
			{Scully}},\ }\bibfield  {title} {\bibinfo {title} {{Enhancing photovoltaic
				power by {Fano}-induced coherence}},\ }\href
	{https://doi.org/10.1103/PhysRevA.84.053818} {\bibfield  {journal} {\bibinfo
			{journal} {Phys. Rev. A}\ }\textbf {\bibinfo {volume} {84}},\ \bibinfo
		{pages} {53818} (\bibinfo {year} {2011})}\BibitemShut {NoStop}%
	\bibitem [{\citenamefont {Aroutiounian}\ \emph {et~al.}(2012)\citenamefont
		{Aroutiounian}, \citenamefont {Petrosyan}, \citenamefont {Khachatryan},\ and\
		\citenamefont {Touryan}}]{Aroutiounian2012}%
	\BibitemOpen
	\bibfield  {author} {\bibinfo {author} {\bibfnamefont {V.}~\bibnamefont
			{Aroutiounian}}, \bibinfo {author} {\bibfnamefont {S.}~\bibnamefont
			{Petrosyan}}, \bibinfo {author} {\bibfnamefont {A.}~\bibnamefont
			{Khachatryan}},\ and\ \bibinfo {author} {\bibfnamefont {K.}~\bibnamefont
			{Touryan}},\ }\bibfield  {title} {\bibinfo {title} {{Quantum dot solar
				cells}},\ }\href {https://doi.org/10.1063/1.1339210} {\bibfield  {journal}
		{\bibinfo  {journal} {Physica E: Low-dimensional Systems and Nanostructures}\
		}\textbf {\bibinfo {volume} {2268}},\ \bibinfo {pages} {115} (\bibinfo {year}
		{2012})}\BibitemShut {NoStop}%
	\bibitem [{\citenamefont {Kamat}(2013)}]{Kamat2013}%
	\BibitemOpen
	\bibfield  {author} {\bibinfo {author} {\bibfnamefont {P.~V.}\ \bibnamefont
			{Kamat}},\ }\bibfield  {title} {\bibinfo {title} {{Quantum dot solar cells.
				The next big thing in photovoltaics}},\ }\href
	{https://doi.org/10.1021/jz400052e} {\bibfield  {journal} {\bibinfo
			{journal} {Journal of Physical Chemistry Letters}\ }\textbf {\bibinfo
			{volume} {4}},\ \bibinfo {pages} {908} (\bibinfo {year} {2013})}\BibitemShut
	{NoStop}%
	\bibitem [{\citenamefont {Guimard}\ \emph {et~al.}(2010)\citenamefont
		{Guimard}, \citenamefont {Morihara}, \citenamefont {Bordel}, \citenamefont
		{Tanabe}, \citenamefont {Wakayama}, \citenamefont {Nishioka},\ and\
		\citenamefont {Arakawa}}]{Guimard2010}%
	\BibitemOpen
	\bibfield  {author} {\bibinfo {author} {\bibfnamefont {D.}~\bibnamefont
			{Guimard}}, \bibinfo {author} {\bibfnamefont {R.}~\bibnamefont {Morihara}},
		\bibinfo {author} {\bibfnamefont {D.}~\bibnamefont {Bordel}}, \bibinfo
		{author} {\bibfnamefont {K.}~\bibnamefont {Tanabe}}, \bibinfo {author}
		{\bibfnamefont {Y.}~\bibnamefont {Wakayama}}, \bibinfo {author}
		{\bibfnamefont {M.}~\bibnamefont {Nishioka}},\ and\ \bibinfo {author}
		{\bibfnamefont {Y.}~\bibnamefont {Arakawa}},\ }\bibfield  {title} {\bibinfo
		{title} {{Fabrication of InAs/GaAs quantum dot solar cells with enhanced
				photocurrent and without degradation of open circuit voltage}},\ }\href
	{https://doi.org/10.1063/1.3427392} {\bibfield  {journal} {\bibinfo
			{journal} {Appl. Phys. Lett.}\ }\textbf {\bibinfo {volume} {96}},\ \bibinfo
		{pages} {203507} (\bibinfo {year} {2010})}\BibitemShut {NoStop}%
	\bibitem [{\citenamefont {Beattie}\ \emph {et~al.}(2017)\citenamefont
		{Beattie}, \citenamefont {See}, \citenamefont {Zoppi}, \citenamefont
		{Ushasree}, \citenamefont {Duchamp}, \citenamefont {Farrer}, \citenamefont
		{Ritchie},\ and\ \citenamefont {Tomi{\'{c}}}}]{Beattie2017}%
	\BibitemOpen
	\bibfield  {author} {\bibinfo {author} {\bibfnamefont {N.~S.}\ \bibnamefont
			{Beattie}}, \bibinfo {author} {\bibfnamefont {P.}~\bibnamefont {See}},
		\bibinfo {author} {\bibfnamefont {G.}~\bibnamefont {Zoppi}}, \bibinfo
		{author} {\bibfnamefont {P.~M.}\ \bibnamefont {Ushasree}}, \bibinfo {author}
		{\bibfnamefont {M.}~\bibnamefont {Duchamp}}, \bibinfo {author} {\bibfnamefont
			{I.}~\bibnamefont {Farrer}}, \bibinfo {author} {\bibfnamefont {D.~A.}\
			\bibnamefont {Ritchie}},\ and\ \bibinfo {author} {\bibfnamefont
			{S.}~\bibnamefont {Tomi{\'{c}}}},\ }\bibfield  {title} {\bibinfo {title}
		{{Quantum Engineering of {InAs/GaAs} Quantum Dot Based Intermediate Band
				Solar Cells}},\ }\href@noop {} {\bibfield  {journal} {\bibinfo  {journal}
			{ACS Photonics}\ }\textbf {\bibinfo {volume} {4}},\ \bibinfo {pages} {2745}
		(\bibinfo {year} {2017})}\BibitemShut {NoStop}%
	\bibitem [{\citenamefont {Pan}\ \emph {et~al.}(2013)\citenamefont {Pan},
		\citenamefont {Zhao}, \citenamefont {Wang}, \citenamefont {Zhang},
		\citenamefont {Feng},\ and\ \citenamefont {Zhong}}]{Pan2013}%
	\BibitemOpen
	\bibfield  {author} {\bibinfo {author} {\bibfnamefont {Z.}~\bibnamefont
			{Pan}}, \bibinfo {author} {\bibfnamefont {K.}~\bibnamefont {Zhao}}, \bibinfo
		{author} {\bibfnamefont {J.}~\bibnamefont {Wang}}, \bibinfo {author}
		{\bibfnamefont {H.}~\bibnamefont {Zhang}}, \bibinfo {author} {\bibfnamefont
			{Y.}~\bibnamefont {Feng}},\ and\ \bibinfo {author} {\bibfnamefont
			{X.}~\bibnamefont {Zhong}},\ }\bibfield  {title} {\bibinfo {title} {Near
			infrared absorption of cdsexte1–x alloyed quantum dot sensitized solar
			cells with more than 6\% efficiency and high stability},\ }\href
	{https://doi.org/10.1021/nn400947e} {\bibfield  {journal} {\bibinfo
			{journal} {ACS Nano}\ }\textbf {\bibinfo {volume} {7}},\ \bibinfo {pages}
		{5215} (\bibinfo {year} {2013})},\ \bibinfo {note} {pMID: 23705771},\ \Eprint
	{https://arxiv.org/abs/https://doi.org/10.1021/nn400947e}
	{https://doi.org/10.1021/nn400947e} \BibitemShut {NoStop}%
	\bibitem [{\citenamefont {Santra}\ and\ \citenamefont
		{Kamat}(2012)}]{Santra2012}%
	\BibitemOpen
	\bibfield  {author} {\bibinfo {author} {\bibfnamefont {P.~K.}\ \bibnamefont
			{Santra}}\ and\ \bibinfo {author} {\bibfnamefont {P.~V.}\ \bibnamefont
			{Kamat}},\ }\bibfield  {title} {\bibinfo {title} {{Mn-Doped Quantum Dot
				Sensitized Solar Cells: A Strategy to Boost Efficiency over 5{\%}}},\ }\href
	{https://doi.org/10.1021/ja211224s} {\bibfield  {journal} {\bibinfo
			{journal} {Journal of the American Chemical Society}\ }\textbf {\bibinfo
			{volume} {134}},\ \bibinfo {pages} {2508} (\bibinfo {year}
		{2012})}\BibitemShut {NoStop}%
	\bibitem [{\citenamefont {Cirloganu}\ \emph {et~al.}(2014)\citenamefont
		{Cirloganu}, \citenamefont {Padilha}, \citenamefont {Lin}, \citenamefont
		{Makarov}, \citenamefont {Velizhanin}, \citenamefont {Luo}, \citenamefont
		{Robel}, \citenamefont {Pietryga},\ and\ \citenamefont
		{Klimov}}]{Cirloganu2014}%
	\BibitemOpen
	\bibfield  {author} {\bibinfo {author} {\bibfnamefont {C.~M.}\ \bibnamefont
			{Cirloganu}}, \bibinfo {author} {\bibfnamefont {L.~A.}\ \bibnamefont
			{Padilha}}, \bibinfo {author} {\bibfnamefont {Q.}~\bibnamefont {Lin}},
		\bibinfo {author} {\bibfnamefont {N.~S.}\ \bibnamefont {Makarov}}, \bibinfo
		{author} {\bibfnamefont {K.~A.}\ \bibnamefont {Velizhanin}}, \bibinfo
		{author} {\bibfnamefont {H.}~\bibnamefont {Luo}}, \bibinfo {author}
		{\bibfnamefont {I.}~\bibnamefont {Robel}}, \bibinfo {author} {\bibfnamefont
			{J.~M.}\ \bibnamefont {Pietryga}},\ and\ \bibinfo {author} {\bibfnamefont
			{V.~I.}\ \bibnamefont {Klimov}},\ }\bibfield  {title} {\bibinfo {title}
		{{Enhanced carrier multiplication in engineered quasi-type-{\{}II{\}} quantum
				dots}},\ }\href {https://doi.org/10.1038/ncomms5148} {\bibfield  {journal}
		{\bibinfo  {journal} {Nature Communications}\ }\textbf {\bibinfo {volume}
			{5}},\ \bibinfo {pages} {4148} (\bibinfo {year} {2014})}\BibitemShut
	{NoStop}%
	\bibitem [{\citenamefont {Lee}\ \emph {et~al.}(2013)\citenamefont {Lee},
		\citenamefont {Son}, \citenamefont {Ahn}, \citenamefont {Shin}, \citenamefont
		{Kim}, \citenamefont {Hwang}, \citenamefont {Ko}, \citenamefont {Sul},
		\citenamefont {Han},\ and\ \citenamefont {Park}}]{Lee2013}%
	\BibitemOpen
	\bibfield  {author} {\bibinfo {author} {\bibfnamefont {J.~W.}\ \bibnamefont
			{Lee}}, \bibinfo {author} {\bibfnamefont {D.~Y.}\ \bibnamefont {Son}},
		\bibinfo {author} {\bibfnamefont {T.~K.}\ \bibnamefont {Ahn}}, \bibinfo
		{author} {\bibfnamefont {H.~W.}\ \bibnamefont {Shin}}, \bibinfo {author}
		{\bibfnamefont {I.~Y.}\ \bibnamefont {Kim}}, \bibinfo {author} {\bibfnamefont
			{S.~J.}\ \bibnamefont {Hwang}}, \bibinfo {author} {\bibfnamefont {M.~J.}\
			\bibnamefont {Ko}}, \bibinfo {author} {\bibfnamefont {S.}~\bibnamefont
			{Sul}}, \bibinfo {author} {\bibfnamefont {H.}~\bibnamefont {Han}},\ and\
		\bibinfo {author} {\bibfnamefont {N.~G.}\ \bibnamefont {Park}},\ }\bibfield
	{title} {\bibinfo {title} {{Quantum-dot-sensitized solar cell with
				unprecedentedly high photocurrent}},\ }\href
	{https://doi.org/10.1038/srep01050} {\bibfield  {journal} {\bibinfo
			{journal} {Scientific Reports}\ }\textbf {\bibinfo {volume} {3}},\ \bibinfo
		{pages} {1} (\bibinfo {year} {2013})}\BibitemShut {NoStop}%
	\bibitem [{\citenamefont {Polman}\ \emph {et~al.}(2016)\citenamefont {Polman},
		\citenamefont {Knight}, \citenamefont {Garnett}, \citenamefont {Ehrler},\
		and\ \citenamefont {Sinke}}]{Polman2016}%
	\BibitemOpen
	\bibfield  {author} {\bibinfo {author} {\bibfnamefont {A.}~\bibnamefont
			{Polman}}, \bibinfo {author} {\bibfnamefont {M.}~\bibnamefont {Knight}},
		\bibinfo {author} {\bibfnamefont {E.~C.}\ \bibnamefont {Garnett}}, \bibinfo
		{author} {\bibfnamefont {B.}~\bibnamefont {Ehrler}},\ and\ \bibinfo {author}
		{\bibfnamefont {W.~C.}\ \bibnamefont {Sinke}},\ }\bibfield  {title} {\bibinfo
		{title} {{Photovoltaic materials: Present efficiencies and future
				challenges}},\ }\href {https://doi.org/10.1126/science.aad4424} {\bibfield
		{journal} {\bibinfo  {journal} {Science}\ }\textbf {\bibinfo {volume}
			{352}},\ \bibinfo {pages} {aad4424} (\bibinfo {year} {2016})}\BibitemShut
	{NoStop}%
	\bibitem [{\citenamefont {Luque}\ and\ \citenamefont
		{Mart\'{i}}(1997)}]{Luque1997}%
	\BibitemOpen
	\bibfield  {author} {\bibinfo {author} {\bibfnamefont {A.}~\bibnamefont
			{Luque}}\ and\ \bibinfo {author} {\bibfnamefont {A.}~\bibnamefont
			{Mart\'{i}}},\ }\bibfield  {title} {\bibinfo {title} {{Increasing the
				Efficiency of Ideal Solar Cells by Photon Induced Transitions at Intermediate
				Levels}},\ }\href@noop {} {\bibfield  {journal} {\bibinfo  {journal} {Phys.
				Rev. Lett.}\ }\textbf {\bibinfo {volume} {78}},\ \bibinfo {pages} {5014}
		(\bibinfo {year} {1997})}\BibitemShut {NoStop}%
	\bibitem [{\citenamefont {Okada}\ \emph {et~al.}(2015)\citenamefont {Okada},
		\citenamefont {Ekins-Daukes}, \citenamefont {Kita}, \citenamefont {Tamaki},
		\citenamefont {Yoshida}, \citenamefont {Pusch}, \citenamefont {Hess},
		\citenamefont {Phillips}, \citenamefont {Farrell}, \citenamefont {Yoshida},
		\citenamefont {Ahsan}, \citenamefont {Shoji}, \citenamefont {Sogabe},\ and\
		\citenamefont {Guillemoles}}]{Okada2015}%
	\BibitemOpen
	\bibfield  {author} {\bibinfo {author} {\bibfnamefont {Y.}~\bibnamefont
			{Okada}}, \bibinfo {author} {\bibfnamefont {N.~J.}\ \bibnamefont
			{Ekins-Daukes}}, \bibinfo {author} {\bibfnamefont {T.}~\bibnamefont {Kita}},
		\bibinfo {author} {\bibfnamefont {R.}~\bibnamefont {Tamaki}}, \bibinfo
		{author} {\bibfnamefont {M.}~\bibnamefont {Yoshida}}, \bibinfo {author}
		{\bibfnamefont {A.}~\bibnamefont {Pusch}}, \bibinfo {author} {\bibfnamefont
			{O.}~\bibnamefont {Hess}}, \bibinfo {author} {\bibfnamefont {C.~C.}\
			\bibnamefont {Phillips}}, \bibinfo {author} {\bibfnamefont {D.~J.}\
			\bibnamefont {Farrell}}, \bibinfo {author} {\bibfnamefont {K.}~\bibnamefont
			{Yoshida}}, \bibinfo {author} {\bibfnamefont {N.}~\bibnamefont {Ahsan}},
		\bibinfo {author} {\bibfnamefont {Y.}~\bibnamefont {Shoji}}, \bibinfo
		{author} {\bibfnamefont {T.}~\bibnamefont {Sogabe}},\ and\ \bibinfo {author}
		{\bibfnamefont {J.~F.}\ \bibnamefont {Guillemoles}},\ }\bibfield  {title}
	{\bibinfo {title} {{Intermediate band solar cells: {\{}Recent{\}} progress
				and future directions}},\ }\href {https://doi.org/10.1063/1.4916561}
	{\bibfield  {journal} {\bibinfo  {journal} {Applied Physics Reviews}\
		}\textbf {\bibinfo {volume} {2}},\ \bibinfo {pages} {21302} (\bibinfo {year}
		{2015})}\BibitemShut {NoStop}%
	\bibitem [{\citenamefont {Bang}\ and\ \citenamefont {Kamat}(2009)}]{Bang2009}%
	\BibitemOpen
	\bibfield  {author} {\bibinfo {author} {\bibfnamefont {J.~H.}\ \bibnamefont
			{Bang}}\ and\ \bibinfo {author} {\bibfnamefont {P.~V.}\ \bibnamefont
			{Kamat}},\ }\bibfield  {title} {\bibinfo {title} {{Quantum Dot Sensitized
				Solar Cells. A Tale of Two Semiconductor Nanocrystals: CdSe and CdTe}},\
	}\href {https://doi.org/10.1021/nn900324q} {\bibfield  {journal} {\bibinfo
			{journal} {ACS Nano}\ }\textbf {\bibinfo {volume} {3}},\ \bibinfo {pages}
		{1467} (\bibinfo {year} {2009})}\BibitemShut {NoStop}%
	\bibitem [{\citenamefont {Bracker}\ \emph {et~al.}(2006)\citenamefont
		{Bracker}, \citenamefont {Scheibner}, \citenamefont {Doty}, \citenamefont
		{Stinaff}, \citenamefont {Ponomarev}, \citenamefont {Kim}, \citenamefont
		{Whitman}, \citenamefont {Reinecke},\ and\ \citenamefont
		{Gammon}}]{Bracker2006}%
	\BibitemOpen
	\bibfield  {author} {\bibinfo {author} {\bibfnamefont {A.~S.}\ \bibnamefont
			{Bracker}}, \bibinfo {author} {\bibfnamefont {M.}~\bibnamefont {Scheibner}},
		\bibinfo {author} {\bibfnamefont {M.~F.}\ \bibnamefont {Doty}}, \bibinfo
		{author} {\bibfnamefont {E.~A.}\ \bibnamefont {Stinaff}}, \bibinfo {author}
		{\bibfnamefont {I.~V.}\ \bibnamefont {Ponomarev}}, \bibinfo {author}
		{\bibfnamefont {J.~C.}\ \bibnamefont {Kim}}, \bibinfo {author} {\bibfnamefont
			{L.~J.}\ \bibnamefont {Whitman}}, \bibinfo {author} {\bibfnamefont {T.~L.}\
			\bibnamefont {Reinecke}},\ and\ \bibinfo {author} {\bibfnamefont
			{D.}~\bibnamefont {Gammon}},\ }\bibfield  {title} {\bibinfo {title}
		{{Engineering electron and hole tunneling with asymmetric InAs quantum dot
				molecules}},\ }\href {https://doi.org/10.1063/1.2400397} {\bibfield
		{journal} {\bibinfo  {journal} {Appl. Phys. Lett.}\ }\textbf {\bibinfo
			{volume} {89}},\ \bibinfo {pages} {233110} (\bibinfo {year}
		{2006})}\BibitemShut {NoStop}%
	\bibitem [{\citenamefont {Ramsay}(2010)}]{Ramsay2010}%
	\BibitemOpen
	\bibfield  {author} {\bibinfo {author} {\bibfnamefont {A.~J.}\ \bibnamefont
			{Ramsay}},\ }\bibfield  {title} {\bibinfo {title} {{A review of the coherent
				optical control of the exciton and spin states of semiconductor quantum
				dots}},\ }\href@noop {} {\bibfield  {journal} {\bibinfo  {journal}
			{Semiconductor Science and Technology}\ }\textbf {\bibinfo {volume} {25}},\
		\bibinfo {pages} {1} (\bibinfo {year} {2010})}\BibitemShut {NoStop}%
	\bibitem [{\citenamefont {Kim}\ \emph {et~al.}(2011)\citenamefont {Kim},
		\citenamefont {Carter}, \citenamefont {Greilich}, \citenamefont {Bracker},\
		and\ \citenamefont {Gammon}}]{Kim2011}%
	\BibitemOpen
	\bibfield  {author} {\bibinfo {author} {\bibfnamefont {D.}~\bibnamefont
			{Kim}}, \bibinfo {author} {\bibfnamefont {S.~G.}\ \bibnamefont {Carter}},
		\bibinfo {author} {\bibfnamefont {A.}~\bibnamefont {Greilich}}, \bibinfo
		{author} {\bibfnamefont {A.~S.}\ \bibnamefont {Bracker}},\ and\ \bibinfo
		{author} {\bibfnamefont {D.}~\bibnamefont {Gammon}},\ }\bibfield  {title}
	{\bibinfo {title} {{Ultrafast optical control of entanglement between two
				quantum dot spins}},\ }\href@noop {} {\bibfield  {journal} {\bibinfo
			{journal} {Nature Physics}\ }\textbf {\bibinfo {volume} {7}},\ \bibinfo
		{pages} {223} (\bibinfo {year} {2011})}\BibitemShut {NoStop}%
	\bibitem [{\citenamefont {Greilich}\ \emph {et~al.}(2011)\citenamefont
		{Greilich}, \citenamefont {Carter}, \citenamefont {Kim}, \citenamefont
		{Bracker},\ and\ \citenamefont {Gammon}}]{Greilich2011}%
	\BibitemOpen
	\bibfield  {author} {\bibinfo {author} {\bibfnamefont {A.}~\bibnamefont
			{Greilich}}, \bibinfo {author} {\bibfnamefont {S.~G.}\ \bibnamefont
			{Carter}}, \bibinfo {author} {\bibfnamefont {D.}~\bibnamefont {Kim}},
		\bibinfo {author} {\bibfnamefont {A.~S.}\ \bibnamefont {Bracker}},\ and\
		\bibinfo {author} {\bibfnamefont {D.}~\bibnamefont {Gammon}},\ }\bibfield
	{title} {\bibinfo {title} {{Optical control of one and two hole spins in
				interacting quantum dots}},\ }\href@noop {} {\bibfield  {journal} {\bibinfo
			{journal} {Nature Photonics}\ }\textbf {\bibinfo {volume} {5}},\ \bibinfo
		{pages} {702} (\bibinfo {year} {2011})}\BibitemShut {NoStop}%
	\bibitem [{\citenamefont {Weiss}\ \emph {et~al.}(2012)\citenamefont {Weiss},
		\citenamefont {Elzerman}, \citenamefont {Delley}, \citenamefont
		{Miguel-Sanchez},\ and\ \citenamefont {Imamo\ifmmode~\breve{g}\else
			\u{g}\fi{}lu}}]{Weiss2012}%
	\BibitemOpen
	\bibfield  {author} {\bibinfo {author} {\bibfnamefont {K.~M.}\ \bibnamefont
			{Weiss}}, \bibinfo {author} {\bibfnamefont {J.~M.}\ \bibnamefont {Elzerman}},
		\bibinfo {author} {\bibfnamefont {Y.~L.}\ \bibnamefont {Delley}}, \bibinfo
		{author} {\bibfnamefont {J.}~\bibnamefont {Miguel-Sanchez}},\ and\ \bibinfo
		{author} {\bibfnamefont {A.}~\bibnamefont {Imamo\ifmmode~\breve{g}\else
				\u{g}\fi{}lu}},\ }\bibfield  {title} {\bibinfo {title} {Coherent two-electron
			spin qubits in an optically active pair of coupled ingaas quantum dots},\
	}\href {https://doi.org/10.1103/PhysRevLett.109.107401} {\bibfield  {journal}
		{\bibinfo  {journal} {Phys. Rev. Lett.}\ }\textbf {\bibinfo {volume} {109}},\
		\bibinfo {pages} {107401} (\bibinfo {year} {2012})}\BibitemShut {NoStop}%
	\bibitem [{\citenamefont {Borges}\ \emph {et~al.}(2012)\citenamefont {Borges},
		\citenamefont {Sanz}, \citenamefont {Villas-B\^oas}, \citenamefont
		{Diniz~Neto},\ and\ \citenamefont {Alcalde}}]{Borges2012}%
	\BibitemOpen
	\bibfield  {author} {\bibinfo {author} {\bibfnamefont {H.~S.}\ \bibnamefont
			{Borges}}, \bibinfo {author} {\bibfnamefont {L.}~\bibnamefont {Sanz}},
		\bibinfo {author} {\bibfnamefont {J.~M.}\ \bibnamefont {Villas-B\^oas}},
		\bibinfo {author} {\bibfnamefont {O.~O.}\ \bibnamefont {Diniz~Neto}},\ and\
		\bibinfo {author} {\bibfnamefont {A.~M.}\ \bibnamefont {Alcalde}},\
	}\bibfield  {title} {\bibinfo {title} {Tunneling induced transparency and
			slow light in quantum dot molecules},\ }\href
	{https://doi.org/10.1103/PhysRevB.85.115425} {\bibfield  {journal} {\bibinfo
			{journal} {Phys. Rev. B}\ }\textbf {\bibinfo {volume} {85}},\ \bibinfo
		{pages} {115425} (\bibinfo {year} {2012})}\BibitemShut {NoStop}%
	\bibitem [{\citenamefont {Borges}\ \emph {et~al.}(2013)\citenamefont {Borges},
		\citenamefont {Sanz}, \citenamefont {Villas-Boas},\ and\ \citenamefont
		{Alcalde}}]{Borges2013}%
	\BibitemOpen
	\bibfield  {author} {\bibinfo {author} {\bibfnamefont {H.~S.}\ \bibnamefont
			{Borges}}, \bibinfo {author} {\bibfnamefont {L.}~\bibnamefont {Sanz}},
		\bibinfo {author} {\bibfnamefont {J.~M.}\ \bibnamefont {Villas-Boas}},\ and\
		\bibinfo {author} {\bibfnamefont {A.~M.}\ \bibnamefont {Alcalde}},\
	}\bibfield  {title} {\bibinfo {title} {{Quantum interference and control of
				the optical response in quantum dot molecules}},\ }\href@noop {} {\bibfield
		{journal} {\bibinfo  {journal} {Appl. Phys. Lett.}\ }\textbf {\bibinfo
			{volume} {103}} (\bibinfo {year} {2013})}\BibitemShut {NoStop}%
	\bibitem [{\citenamefont {Chia}\ \emph {et~al.}(2007)\citenamefont {Chia},
		\citenamefont {Zhang}, \citenamefont {Wong}, \citenamefont {Yong},
		\citenamefont {Chow}, \citenamefont {Chua},\ and\ \citenamefont
		{Guo}}]{Chia07}%
	\BibitemOpen
	\bibfield  {author} {\bibinfo {author} {\bibfnamefont {C.~K.}\ \bibnamefont
			{Chia}}, \bibinfo {author} {\bibfnamefont {Y.~W.}\ \bibnamefont {Zhang}},
		\bibinfo {author} {\bibfnamefont {S.~S.}\ \bibnamefont {Wong}}, \bibinfo
		{author} {\bibfnamefont {A.~M.}\ \bibnamefont {Yong}}, \bibinfo {author}
		{\bibfnamefont {S.~Y.}\ \bibnamefont {Chow}}, \bibinfo {author}
		{\bibfnamefont {S.~J.}\ \bibnamefont {Chua}},\ and\ \bibinfo {author}
		{\bibfnamefont {J.}~\bibnamefont {Guo}},\ }\bibfield  {title} {\bibinfo
		{title} {{Saturated dot density of {InAs/GaAs} self-assembled quantum dots
				grown at high growth rate}},\ }\href {https://doi.org/10.1063/1.2724776}
	{\bibfield  {journal} {\bibinfo  {journal} {Appl. Phys. Lett.}\ }\textbf
		{\bibinfo {volume} {90}},\ \bibinfo {pages} {161906} (\bibinfo {year}
		{2007})}\BibitemShut {NoStop}%
	\bibitem [{Note1()}]{Note1}%
	\BibitemOpen
	\bibinfo {note} {From data of Ref. \cite {Bracker2006}, the electron and hole
		anticrossings energies, $\Delta _{e,h}$, due to tunneling coupling, decrease
		exponentially with $d$. We found by fitting $\Delta _e \simeq 11.67
		e^{-d/7.14}$ and $\Delta _h \simeq 2.2 e^{-d/3.37}$, respectively. Using a
		simple two-level model the corresponding tunneling rates are given by $T_i =
		\Delta _i /2\hbar $, with $i=e,h$.}\BibitemShut {Stop}%
	\bibitem [{Note2()}]{Note2}%
	\BibitemOpen
	\bibinfo {note} {Although manipulations on the growth process could enhance
		the coupling between electrons and phonons for some very specific situations,
		the coupling can be considered weak, because the local charge inside the
		quantum dot is neutral (each dot has the same number of electrons and holes
		in the steady-state).}\BibitemShut {Stop}%
	\bibitem [{\citenamefont {Heitz}\ \emph {et~al.}(2001)\citenamefont {Heitz},
		\citenamefont {Born}, \citenamefont {Guffarth}, \citenamefont {Stier},
		\citenamefont {Schliwa}, \citenamefont {Hoffmann},\ and\ \citenamefont
		{Bimberg}}]{Heitz01}%
	\BibitemOpen
	\bibfield  {author} {\bibinfo {author} {\bibfnamefont {R.}~\bibnamefont
			{Heitz}}, \bibinfo {author} {\bibfnamefont {H.}~\bibnamefont {Born}},
		\bibinfo {author} {\bibfnamefont {F.}~\bibnamefont {Guffarth}}, \bibinfo
		{author} {\bibfnamefont {O.}~\bibnamefont {Stier}}, \bibinfo {author}
		{\bibfnamefont {A.}~\bibnamefont {Schliwa}}, \bibinfo {author} {\bibfnamefont
			{A.}~\bibnamefont {Hoffmann}},\ and\ \bibinfo {author} {\bibfnamefont
			{D.}~\bibnamefont {Bimberg}},\ }\bibfield  {title} {\bibinfo {title}
		{Existence of a phonon bottleneck for excitons in quantum dots},\ }\href
	{https://doi.org/10.1103/PhysRevB.64.241305} {\bibfield  {journal} {\bibinfo
			{journal} {Phys. Rev. B}\ }\textbf {\bibinfo {volume} {64}},\ \bibinfo
		{pages} {241305} (\bibinfo {year} {2001})}\BibitemShut {NoStop}%
	\bibitem [{\citenamefont {Schmitt-Rink}\ \emph {et~al.}(1987)\citenamefont
		{Schmitt-Rink}, \citenamefont {Miller},\ and\ \citenamefont
		{Chemla}}]{Schmitt87}%
	\BibitemOpen
	\bibfield  {author} {\bibinfo {author} {\bibfnamefont {S.}~\bibnamefont
			{Schmitt-Rink}}, \bibinfo {author} {\bibfnamefont {D.~A.~B.}\ \bibnamefont
			{Miller}},\ and\ \bibinfo {author} {\bibfnamefont {D.~S.}\ \bibnamefont
			{Chemla}},\ }\bibfield  {title} {\bibinfo {title} {Theory of the linear and
			nonlinear optical properties of semiconductor microcrystallites},\ }\href
	{https://doi.org/10.1103/PhysRevB.35.8113} {\bibfield  {journal} {\bibinfo
			{journal} {Phys. Rev. B}\ }\textbf {\bibinfo {volume} {35}},\ \bibinfo
		{pages} {8113} (\bibinfo {year} {1987})}\BibitemShut {NoStop}%
	\bibitem [{\citenamefont {Haug}\ and\ \citenamefont {Jauho}(2008)}]{JauhoBook}%
	\BibitemOpen
	\bibfield  {author} {\bibinfo {author} {\bibfnamefont {H.}~\bibnamefont
			{Haug}}\ and\ \bibinfo {author} {\bibfnamefont {A.-P.}\ \bibnamefont
			{Jauho}},\ }\href@noop {} {\emph {\bibinfo {title} {{Quantum Kinetic in
					Transport and Optics of Semiconductors}}}}\ (\bibinfo  {publisher}
	{Springer},\ \bibinfo {address} {Berlin},\ \bibinfo {year}
	{2008})\BibitemShut {NoStop}%
	\bibitem [{\citenamefont {Stinaff}\ \emph {et~al.}(2006)\citenamefont
		{Stinaff}, \citenamefont {Scheibner}, \citenamefont {Bracker}, \citenamefont
		{Ponomarev}, \citenamefont {Korenev}, \citenamefont {Ware}, \citenamefont
		{Doty}, \citenamefont {Reinecke},\ and\ \citenamefont {Gammon}}]{Stinaff636}%
	\BibitemOpen
	\bibfield  {author} {\bibinfo {author} {\bibfnamefont {E.~A.}\ \bibnamefont
			{Stinaff}}, \bibinfo {author} {\bibfnamefont {M.}~\bibnamefont {Scheibner}},
		\bibinfo {author} {\bibfnamefont {A.~S.}\ \bibnamefont {Bracker}}, \bibinfo
		{author} {\bibfnamefont {I.~V.}\ \bibnamefont {Ponomarev}}, \bibinfo {author}
		{\bibfnamefont {V.~L.}\ \bibnamefont {Korenev}}, \bibinfo {author}
		{\bibfnamefont {M.~E.}\ \bibnamefont {Ware}}, \bibinfo {author}
		{\bibfnamefont {M.~F.}\ \bibnamefont {Doty}}, \bibinfo {author}
		{\bibfnamefont {T.~L.}\ \bibnamefont {Reinecke}},\ and\ \bibinfo {author}
		{\bibfnamefont {D.}~\bibnamefont {Gammon}},\ }\bibfield  {title} {\bibinfo
		{title} {{Optical Signatures of Coupled Quantum Dots}},\ }\href@noop {}
	{\bibfield  {journal} {\bibinfo  {journal} {Science}\ }\textbf {\bibinfo
			{volume} {311}},\ \bibinfo {pages} {636} (\bibinfo {year}
		{2006})}\BibitemShut {NoStop}%
	\bibitem [{Note3()}]{Note3}%
	\BibitemOpen
	\bibinfo {note} {This value was obtained after extracting the experimental
		data from Fig. 2 of the external quantum efficiency reported in Guimard
		\protect \textit {et al.} and additional calculations.}\BibitemShut {Stop}%
	\bibitem [{\citenamefont {Fry}\ \emph {et~al.}(2000)\citenamefont {Fry},
		\citenamefont {Itskevich}, \citenamefont {Parnell}, \citenamefont {Finley},
		\citenamefont {Wilson}, \citenamefont {Schumacher}, \citenamefont {Mowbray},
		\citenamefont {Skolnick}, \citenamefont {{Al-Khafaji}}, \citenamefont
		{Cullis}, \citenamefont {Hopkinson}, \citenamefont {Clark},\ and\
		\citenamefont {Hill}}]{Fry2000}%
	\BibitemOpen
	\bibfield  {author} {\bibinfo {author} {\bibfnamefont {P.~W.}\ \bibnamefont
			{Fry}}, \bibinfo {author} {\bibfnamefont {I.~E.}\ \bibnamefont {Itskevich}},
		\bibinfo {author} {\bibfnamefont {S.~R.}\ \bibnamefont {Parnell}}, \bibinfo
		{author} {\bibfnamefont {J.~J.}\ \bibnamefont {Finley}}, \bibinfo {author}
		{\bibfnamefont {L.~R.}\ \bibnamefont {Wilson}}, \bibinfo {author}
		{\bibfnamefont {K.~L.}\ \bibnamefont {Schumacher}}, \bibinfo {author}
		{\bibfnamefont {D.~J.}\ \bibnamefont {Mowbray}}, \bibinfo {author}
		{\bibfnamefont {M.~S.}\ \bibnamefont {Skolnick}}, \bibinfo {author}
		{\bibfnamefont {M.}~\bibnamefont {{Al-Khafaji}}}, \bibinfo {author}
		{\bibfnamefont {A.~G.}\ \bibnamefont {Cullis}}, \bibinfo {author}
		{\bibfnamefont {M.}~\bibnamefont {Hopkinson}}, \bibinfo {author}
		{\bibfnamefont {J.~C.}\ \bibnamefont {Clark}},\ and\ \bibinfo {author}
		{\bibfnamefont {G.}~\bibnamefont {Hill}},\ }\bibfield  {title} {\bibinfo
		{title} {{Photocurrent Spectroscopy of {{InAs}}/{{GaAs}} Self-Assembled
				Quantum Dots}},\ }\href {https://doi.org/10.1103/PhysRevB.62.16784}
	{\bibfield  {journal} {\bibinfo  {journal} {Phys. Rev. B}\ }\textbf {\bibinfo
			{volume} {62}},\ \bibinfo {pages} {16784} (\bibinfo {year}
		{2000})}\BibitemShut {NoStop}%
	\bibitem [{\citenamefont {Brunkov}\ \emph {et~al.}(2002)\citenamefont
		{Brunkov}, \citenamefont {Patan\`e}, \citenamefont {Levin}, \citenamefont
		{Eaves}, \citenamefont {Main}, \citenamefont {Musikhin}, \citenamefont
		{Volovik}, \citenamefont {Zhukov}, \citenamefont {Ustinov},\ and\
		\citenamefont {Konnikov}}]{Brunkov2002}%
	\BibitemOpen
	\bibfield  {author} {\bibinfo {author} {\bibfnamefont {P.~N.}\ \bibnamefont
			{Brunkov}}, \bibinfo {author} {\bibfnamefont {A.}~\bibnamefont {Patan\`e}},
		\bibinfo {author} {\bibfnamefont {A.}~\bibnamefont {Levin}}, \bibinfo
		{author} {\bibfnamefont {L.}~\bibnamefont {Eaves}}, \bibinfo {author}
		{\bibfnamefont {P.~C.}\ \bibnamefont {Main}}, \bibinfo {author}
		{\bibfnamefont {Y.~G.}\ \bibnamefont {Musikhin}}, \bibinfo {author}
		{\bibfnamefont {B.~V.}\ \bibnamefont {Volovik}}, \bibinfo {author}
		{\bibfnamefont {A.~E.}\ \bibnamefont {Zhukov}}, \bibinfo {author}
		{\bibfnamefont {V.~M.}\ \bibnamefont {Ustinov}},\ and\ \bibinfo {author}
		{\bibfnamefont {S.~G.}\ \bibnamefont {Konnikov}},\ }\bibfield  {title}
	{\bibinfo {title} {Photocurrent and capacitance spectroscopy of schottky
			barrier structures incorporating {InAs/GaAs} quantum dots},\ }\href
	{https://doi.org/10.1103/PhysRevB.65.085326} {\bibfield  {journal} {\bibinfo
			{journal} {Phys. Rev. B}\ }\textbf {\bibinfo {volume} {65}},\ \bibinfo
		{pages} {085326} (\bibinfo {year} {2002})}\BibitemShut {NoStop}%
	\bibitem [{Note4()}]{Note4}%
	\BibitemOpen
	\bibinfo {note} {In a TLS model with phenomenological population $\gamma _0$
		and coherence $\gamma ^\prime $ damping rates, the steady-state population
		and coherence are given by $\rho _{ee} = \protect \frac {W^2}{2\gamma _0
			\gamma ^\prime } \protect \frac {1}{1 + \protect \frac {\delta ^2}{{\gamma
					^\prime }^2} + \protect \frac {W^2}{\gamma _0 \gamma ^\prime } }$, $\rho
		_{eg} = -i \protect \frac {W}{2 \gamma ^\prime } \protect \frac {1 + \protect
			\frac {i\delta }{\gamma ^\prime }}{1 + \protect \frac {\delta ^2 }{{\gamma
					^\prime }^2} + \protect \frac {W^2}{\gamma _0 \gamma ^\prime }}$,
		respectively.}\BibitemShut {Stop}%
	\bibitem [{\citenamefont {Gawarecki}\ \emph {et~al.}(2010)\citenamefont
		{Gawarecki}, \citenamefont {Pochwa\l{}a}, \citenamefont {Grodecka-Grad},\
		and\ \citenamefont {Machnikowski}}]{Gawarecki10}%
	\BibitemOpen
	\bibfield  {author} {\bibinfo {author} {\bibfnamefont {K.}~\bibnamefont
			{Gawarecki}}, \bibinfo {author} {\bibfnamefont {M.}~\bibnamefont
			{Pochwa\l{}a}}, \bibinfo {author} {\bibfnamefont {A.}~\bibnamefont
			{Grodecka-Grad}},\ and\ \bibinfo {author} {\bibfnamefont {P.}~\bibnamefont
			{Machnikowski}},\ }\bibfield  {title} {\bibinfo {title} {Phonon-assisted
			relaxation and tunneling in self-assembled quantum dot molecules},\
	}\href@noop {} {\bibfield  {journal} {\bibinfo  {journal} {Phys. Rev. B}\
		}\textbf {\bibinfo {volume} {81}},\ \bibinfo {pages} {245312} (\bibinfo
		{year} {2010})}\BibitemShut {NoStop}%
	\bibitem [{\citenamefont {M\"uller}\ \emph {et~al.}(2012)\citenamefont
		{M\"uller}, \citenamefont {Bechtold}, \citenamefont {Ruppert}, \citenamefont
		{Zecherle}, \citenamefont {Reithmaier}, \citenamefont {Bichler},
		\citenamefont {Krenner}, \citenamefont {Abstreiter}, \citenamefont
		{Holleitner}, \citenamefont {Villas-Boas}, \citenamefont {Betz},\ and\
		\citenamefont {Finley}}]{Muller12}%
	\BibitemOpen
	\bibfield  {author} {\bibinfo {author} {\bibfnamefont {K.}~\bibnamefont
			{M\"uller}}, \bibinfo {author} {\bibfnamefont {A.}~\bibnamefont {Bechtold}},
		\bibinfo {author} {\bibfnamefont {C.}~\bibnamefont {Ruppert}}, \bibinfo
		{author} {\bibfnamefont {M.}~\bibnamefont {Zecherle}}, \bibinfo {author}
		{\bibfnamefont {G.}~\bibnamefont {Reithmaier}}, \bibinfo {author}
		{\bibfnamefont {M.}~\bibnamefont {Bichler}}, \bibinfo {author} {\bibfnamefont
			{H.~J.}\ \bibnamefont {Krenner}}, \bibinfo {author} {\bibfnamefont
			{G.}~\bibnamefont {Abstreiter}}, \bibinfo {author} {\bibfnamefont {A.~W.}\
			\bibnamefont {Holleitner}}, \bibinfo {author} {\bibfnamefont {J.~M.}\
			\bibnamefont {Villas-Boas}}, \bibinfo {author} {\bibfnamefont
			{M.}~\bibnamefont {Betz}},\ and\ \bibinfo {author} {\bibfnamefont {J.~J.}\
			\bibnamefont {Finley}},\ }\bibfield  {title} {\bibinfo {title} {Electrical
			control of interdot electron tunneling in a double {InGaAs} quantum-dot
			nanostructure},\ }\href@noop {} {\bibfield  {journal} {\bibinfo  {journal}
			{Phys. Rev. Lett.}\ }\textbf {\bibinfo {volume} {108}},\ \bibinfo {pages}
		{197402} (\bibinfo {year} {2012})}\BibitemShut {NoStop}%
	\bibitem [{\citenamefont {Nakaoka}\ \emph {et~al.}(2006)\citenamefont
		{Nakaoka}, \citenamefont {Clark}, \citenamefont {Krenner}, \citenamefont
		{Sabathil}, \citenamefont {Bichler}, \citenamefont {Arakawa}, \citenamefont
		{Abstreiter},\ and\ \citenamefont {Finley}}]{Nakaoka06}%
	\BibitemOpen
	\bibfield  {author} {\bibinfo {author} {\bibfnamefont {T.}~\bibnamefont
			{Nakaoka}}, \bibinfo {author} {\bibfnamefont {E.~C.}\ \bibnamefont {Clark}},
		\bibinfo {author} {\bibfnamefont {H.~J.}\ \bibnamefont {Krenner}}, \bibinfo
		{author} {\bibfnamefont {M.}~\bibnamefont {Sabathil}}, \bibinfo {author}
		{\bibfnamefont {M.}~\bibnamefont {Bichler}}, \bibinfo {author} {\bibfnamefont
			{Y.}~\bibnamefont {Arakawa}}, \bibinfo {author} {\bibfnamefont
			{G.}~\bibnamefont {Abstreiter}},\ and\ \bibinfo {author} {\bibfnamefont
			{J.~J.}\ \bibnamefont {Finley}},\ }\bibfield  {title} {\bibinfo {title}
		{Direct observation of acoustic phonon mediated relaxation between coupled
			exciton states in a single quantum dot molecule},\ }\href@noop {} {\bibfield
		{journal} {\bibinfo  {journal} {Phys. Rev. B}\ }\textbf {\bibinfo {volume}
			{74}},\ \bibinfo {pages} {121305} (\bibinfo {year} {2006})}\BibitemShut
	{NoStop}%
	\bibitem [{\citenamefont {Stace}\ \emph {et~al.}(2005)\citenamefont {Stace},
		\citenamefont {Doherty},\ and\ \citenamefont {Barrett}}]{Stace05}%
	\BibitemOpen
	\bibfield  {author} {\bibinfo {author} {\bibfnamefont {T.~M.}\ \bibnamefont
			{Stace}}, \bibinfo {author} {\bibfnamefont {A.~C.}\ \bibnamefont {Doherty}},\
		and\ \bibinfo {author} {\bibfnamefont {S.~D.}\ \bibnamefont {Barrett}},\
	}\bibfield  {title} {\bibinfo {title} {Erratum: Population inversion of a
			driven two-level system in a structureless bath {[Phys. Rev. Lett. 95, 106801
				(2005)]}},\ }\href@noop {} {\bibfield  {journal} {\bibinfo  {journal} {Phys.
				Rev. Lett.}\ }\textbf {\bibinfo {volume} {95}},\ \bibinfo {pages} {209902}
		(\bibinfo {year} {2005})}\BibitemShut {NoStop}%
	\bibitem [{\citenamefont {Gauger}\ \emph {et~al.}(2008)\citenamefont {Gauger},
		\citenamefont {Nazir}, \citenamefont {Stace},\ and\ \citenamefont
		{Lovett}}]{Gauger08}%
	\BibitemOpen
	\bibfield  {author} {\bibinfo {author} {\bibfnamefont {E.~M.}\ \bibnamefont
			{Gauger}}, \bibinfo {author} {\bibfnamefont {S.~C.}\ \bibnamefont {Nazir},
			\bibfnamefont {Ahsan amd~Benjamin}}, \bibinfo {author} {\bibfnamefont
			{T.~M.}\ \bibnamefont {Stace}},\ and\ \bibinfo {author} {\bibfnamefont
			{B.~W.}\ \bibnamefont {Lovett}},\ }\bibfield  {title} {\bibinfo {title}
		{Robust adiabatic approach to optical spin entangling in coupled quantum
			dots},\ }\href@noop {} {\bibfield  {journal} {\bibinfo  {journal} {New J.
				Phys.}\ }\textbf {\bibinfo {volume} {10}},\ \bibinfo {pages} {073016}
		(\bibinfo {year} {2008})}\BibitemShut {NoStop}%
\end{thebibliography}
\end{document}